\PassOptionsToPackage{table,dvipsnames}{xcolor}
\documentclass{article}

     \usepackage[final,nonatbib]{mlrsa_2020}

\usepackage[utf8]{inputenc} 
\usepackage[T1]{fontenc}    
\usepackage{hyperref}       
\usepackage{url}            
\usepackage{booktabs}       
\usepackage{amsfonts}       
\usepackage{nicefrac}       
\usepackage{microtype}      

\usepackage[utf8]{inputenc}
\usepackage[T1]{fontenc}
\usepackage{textcomp}

\usepackage{pgfplotstable}
\usepackage{pgfplots}
\usepackage{amsmath,empheq, cases}
\usepackage{amssymb}
\usepackage{bbm}
\usepackage[frozencache,cachedir=.]{minted}
\usepackage{float}
\usepackage{microtype}
\usepackage{tikz}
\usepackage{forest}
\usetikzlibrary{fit,positioning,arrows,automata,calc}
\usetikzlibrary{shapes,arrows,arrows.meta, positioning,shapes.misc,decorations.markings}
\usepackage{booktabs}
\usepackage{multirow}
\usepackage{adjustbox}
\usepackage{verbatim}
\usepackage[title]{appendix}
\usepackage[font={small,singlespacing},labelfont=it,labelsep=period]{caption}
\usepackage{adjustbox} 
\usepackage{graphicx}
\usepackage{subcaption}
\usepackage{natbib}

\newcommand\textcite[1]{\citet{#1}}
\renewcommand\cite[1]{\citep{#1}}
\usepackage[autostyle, english=american]{csquotes}
\MakeOuterQuote{"}

\usepackage{neuralnetwork}

\usepackage[colorinlistoftodos,disable]{todonotes}
\usepackage{newunicodechar}
\newunicodechar{ė}{\.{e}}
\newunicodechar{Ė}{\.{E}}
\newcommand{\norm}[1]{\left\lVert#1\right\rVert}
\DeclareMathOperator{\sign}{sign}

\newcolumntype{t}{>{\ttfamily}l}
\newcolumntype{T}{>{\ttfamily}c}

\newcolumntype{$}{>{\global\let\currentrowstyle\relax}}
\newcolumntype{^}{>{\currentrowstyle}}

\title{A Survey of Machine Learning Methods and Challenges for Windows Malware Classification}

\author{Edward Raff\\
Booz Allen Hamilton\\
  \texttt{raff\_edward@bah.com} \\
  University of Maryland, Baltimore County\\
  \texttt{raff.edward@umbc.edu} \\
\And Charles Nicholas\\ University of Maryland, Baltimore County 
  \\ \texttt{nicholas@umbc.edu}
}

\begin{document}

\maketitle

\begin{abstract}
Malware classification is a difficult problem, to which machine learning methods have been applied for decades. Yet progress has often been slow, in part due to a number of unique difficulties with the task that occur through all stages of the developing a machine learning system: data collection, labeling, feature creation and selection, model selection, and evaluation. In this survey we will review a number of the current methods and challenges related to malware classification, including data collection, feature extraction, and model construction, and evaluation. Our discussion will include thoughts on the constraints that must be considered for machine learning based solutions in this domain, and yet to be tackled problems for which machine learning could also provide a solution. This survey aims to be useful both to cybersecurity practitioners who wish to learn more about how machine learning can be applied to the malware problem, and to give data scientists the necessary background into the challenges in this uniquely complicated space. 
\end{abstract}

\section{Introduction} \label{sec:intro}

The impact of malicious software, or "malware", such as viruses and worms, is a long standing and growing problem. As society becomes increasingly dependent on computer systems, this impact only increases. Single incidents regularly cost companies tens of millions of dollars in damages \cite{Musil2016,Riley2015,Frizell2015}. In 2014, for example,
the total economic cost associated with malware distributed through pirated software, a subset of all malware, was estimated to be nearly \$500 billion \cite{Gantz2014}. Overall malware has proven to be effective for its authors, as indicated by the exponential growth of new malware \cite{tagkey2014iv,AV-TEST2016,deepguard_F-Secure2016}. This growth in malware only increases the need for better tools to stop malware and aid analysts and security professionals. 

One specific area for improvement is \textit{malware classification}. The task of malware classification has been long studied, and generally refers to one of two related tasks: 1) detecting new malware (i.e., distinguishing between benign and malicious applications) and 2) differentiating between two or more known malware types or families. The former of these we will refer to as \textit{malware detection}, and it is intrinsically useful in stopping the spread of malware. Anti-virus (AV) products currently perform this function using a predominantly signature-based approach. Signatures are intrinsically specific to the malware they detect, and can be 
labor-intensive for an analyst to create. This makes signatures unlikely to scale as malware becomes more prevalent, an issue publicly recognized by AV vendors \cite{Yadron2014,Hypponen2012}. 

The second class of malware classification we will refer to as \textit{malware family classification}. Analysts and security professionals use this process to sort through new binaries and process an ever growing amount of data. In this case we assume or know that the binary is malicious, and wish to designate it as a member of a specific family. For example, Conficker is a specific malware family that was prevalent from 2008 through 2009, and evolved through five major revisions \cite{Porras:2009:IRR:1562764.1562777}. Modern obfuscation techniques employed by malware authors means that a single variant of a piece of malware may have several instantiations in the wild that do not have the same MD5 or SHA hash, but do have substantially identical functionality. Thus there is a need to automatically determine if newly observed binaries in fact belong to a previously known family. Such a finding aids in attribution, and reduces the number of files that analysts must look at. 

The field of machine learning would seem the first and most likely avenue to provide relief to the malware classification problem, and indeed ML has been applied to this problem for decades \cite{Kephart:1995:BID:1625855.1625983}. While the problem of malware detection is an issue that spans across many file formats \cite{tabish2009malware} and operating systems, we will focus on the case of Microsoft Portable Executable (PE) binaries for 32 and 64-bit versions of the Windows OS, as was first considered by \textcite{Schultz2001}. In many cases the same issues and methods have been applied to other malware domains, such as malicious PDF files, or malware for the Linux or Android platforms. 
We focus on Windows executables because it is one of the longest standing platforms, and thus has the longest history of continuous research. 
Machine learning is an important contributor to a solution due to its focus on generalization, meaning that the models learned should work effectively on new malware specimens, which have not previously been seen. The importance of having methods that generalize is only growing, as a recent study has found that 94\% of applications on a sample of 100 million machines are unique to the machine \cite{Li2017}. This means a system deployed widely will have to deal with hundreds of millions of unique files that would have never been encountered before. 
The combination of machine learning and malware 
draws from many branches of Computer Science: From high level algorithms, statistics, and information theory to build classification models, to low level details on assembly instruction set architectures, software design, and networking needed to understand how malware works and propagates. Challenges from across this spectrum interact to make malware classification one of the more difficult tasks to which one could apply machine learning. Given that researchers often (if not always) have limited resources and time, it is not surprising that this area has received less attention and struggled to make progress, even as machine learning becomes an increasingly mainstream tool. 
In particular we note that a wide array of potential machine learning methods have not yet been explored for malware classification. Thus in this survey we make an effort to address techniques we feel should be investigated, especially if they are particularly appropriate or readily available in free software projects. 

Our primary goal is to provide a common base of understanding for both malware researchers and machine learning practitioners who are interested in this combined space. What techniques have been looked at, what are their strengths or challenges, and what areas are in need of continued work? 
We present the information about what has been done and why effective application is a challenge from the bottom up, starting with the difficulties of getting representative data in \autoref{sec:data}. This is a critical issue that has been overlooked in many prior works, as all subsequent stages depend on that data being representative of the population at large. Next we will discuss the features that have been extracted from these binaries in \autoref{sec:features}, focusing on both static and dynamic features and their respective strengths and weaknesses. The next three sections will be on the types of machine learning approaches that have been applied to these features. More "standard" machine learning methods used on simple feature vectors are discussed in \autoref{sec:ml_methods}. In \autoref{sec:ml_seq} we will discuss the machine learning models used that understand the input representation as being a type of sequence, which maps better to the true nature of the data. The most expressive representational form, that of a graph, is discussed in \autoref{sec:graphs}.  While this form is one that often best encodes the nature of the features used, it is also the least frequently used. Once we have chosen our data, features, and model, the next step is evaluation in \autoref{sec:evaluation}. Finally, we discuss some future research directions in \autoref{sec:future_research} and our conclusions in \autoref{sec:conclusion}.  In each section of this survey we will attempt to provide a broad overview of what is currently done and the challenges that are impeding progress or need to be considered. 

It is unfortunate that for many of the issues we will discuss, there is little quantitative information about how widespread or significant their impact is.  The complexity of malware ranges from simple applications relying on user error and unpatched vulnerabilities to work, to viruses like Stuxnet - written (it is said) by nation state actors, 
which attempt to evade detection and may have selective intent that goes unnoticed or unused on most systems \cite{Kushner2013}. This broad spectrum of sophistication means that different issues and countermeasures may have a more or less noticeable impact on a learning system depending on the current prevalence of such measures, the malware we would like to classify, and the systems on which a solution would be deployed. This can change over time and it is not currently feasible to tackle all of these issues at once. For these reasons we refrain from declaring any method the "state of the art" for malware classification, and instead focus on the pros and cons of the various approaches, as well as the underlying issues that cause this slippery situation. In particular, we will focus on any theoretical shortcomings that would prevent a system from working in practice, such as, for example, any machine learning processes which an adversary could circumvent with minimal effort if they wished to do so. 

\section{Data Collection Challenges} \label{sec:data}

As with many applications, the first task to building a machine learning model is to obtain data that accurately represents the distribution of binaries that will be observed. It is indeed well known that obtaining more and better labeled data is one of the most effective ways to improve the accuracy of a machine learning system \cite{Domingos:2012:FUT:2347736.2347755,halevy2009unreasonable}. However, by its very nature the potential scope of what a binary can do is unbounded. There is no way for us to randomly sample from the binaries that may exist in the world, and we have no way to measure how much of the "space" of binaries we have covered with any given dataset. 
Beyond the unbounded scope, the malware domain poses a number of unique challenges to data collection. This makes it almost impossible to perform canonical best practices, such as having multiple labelers per file and judging inter-labeler agreement \cite{10.1145/3351095.3372862}. 

When obtaining data, it is often the case that malware is the easiest to get. Not only are there websites dedicated to collecting malware sent in by volunteers \cite{VirusShare,Quist}, but it is not unusual for a researcher to obtain their own malware specimens through the use of honeypots \cite{Baecher2006}. A honeypot is a system connected to the Internet that intentionally tries to get infected by malware, often by leaving open security holes and foregoing standard protections. At the same time, both of these sources of malware can have data quality issues. 
Honeypots will have data biased toward what that system is capable of collecting, as malware may require interaction from the honeypot through specific applications in order to successfully infect a machine \cite{Zhuge2007}.  That is, a malware sample's infection vector may rely on a specific version of Firefox or Chrome to be running, and it may not be possible to account for all possible application interactions. Malware may also attempt to detect that a potential target is in fact a honeypot, and avoid infection to defer its detection \cite{Krawetz2004}. The issues that bias what malware is collected by honeypots are also likely to impact the quality of larger malware repositories, as users may run honeypots and submit their catches to these larger collections. Malware repositories will also have a self-selection bias from those who are willing to share their malware and take the time to do so. 

Benign data, or "goodware", has proven to be even more challenging to physically obtain than malware. This is in part because malware actively attempts to infect new hosts, whereas benign applications do not generally spread prolifically. As far as we are aware, no work has been done to quantify the diversity or collection of benign samples, or how to best obtain representative benign data. Most works take the easiest avenue of data collection, which is to simply collect the binaries found on an installation of Microsoft Windows. This tactic can lead to extreme over-fitting, where models literally learn to find the string "Microsoft Windows" to make a determination \cite{seymor_kaggle_overfit,raff_ngram_2016}. 
The population of binaries from Windows share too much of a common base to 
be useful for training more general models
Instead, the model learns to classify everything that does not come from Microsoft as malware \cite{seymor_kaggle_overfit}. This bias is strong enough that even using only a subset of the information will still lead to over-fitting \cite{raff2017peheader}. This issue is particularly wide spread, and occurs in almost all cited papers in this survey. The significant exception to this are papers produced by corporate entities that have private data they use to develop anti-virus software. When this goodware bias issue is combined with the fact that there is no standard data-set for the task of malware detection, it is almost impossible to compare the results from different papers when different datasets are used. In addition, prior work using benign samples from clean Microsoft installations may significantly over-estimate the accuracy of their methods.

Only recently has effort been made to address this lack of a standard dataset for malware detection.  \textcite{Anderson2018} released the EMBER dataset, which contains features extracted from 1.1 million benign and malicious binaries. EMBER is the first standardized corpus that has been released for malware detection. Their work has taken important steps toward reproducible science and a shared benchmark, but more work is still needed. By the author's own admission, the method of its labeling makes it an "easy" corpus. If users want to create new features from the raw binaries, they have to obtain the binaries themselves independently --- as the authors are unable to share the raw files due to copyright concerns. Information regarding malware family is also not present in the original version. A 2018 version of the Ember corpus (released in 2019, so that labels would be of higher confidence) has attempted to rectify a number of these issues by using more challenging data and malware family information.  

Once data has been obtained, labeling the data must follow (when labels do not come "for free" as they do with honeypots). The issue of labeling malware into families, or determining if an unknown binary is or is not malware, is labor intensive and requires significant domain knowledge and training. This is in contrast to many current machine learning domains, like image classification, where labeling can often be done by individuals with no special expertise and with minimal time. For example, an expert analyst can often take around 10 hours to characterize a malicious binary \cite{Mohaisen:2013:UZA:2487788.2488056}. This observation of the expense to understand what a file does is not unique, with a recent survey reporting hours to weeks of time, with participants ranging from 3-21 years experience \cite{10.1145/3290607.3313040}.This effort makes manually labeling large corpora impractical.
For an entirely expert-labeled corpus for malware family classification, the largest public corpus we are aware of was developed by \textcite{Upchurch2015}. They grouped 85 malware samples by functional similarity into a total of 8 groups. 

For benign vs malicious labeling, many have attempted to circumvent this issue through the use of anti-virus (AV) products. One popular strategy is to upload the binaries to websites such as \textit{VirusTotal}, which will run several dozen AV products against each binary, and return individual results. If more than 30\% of the AVs claim a file is malicious, it is assumed malicious. 
If none of the AVs say it is malware, it is assumed benign. Any specimen that tests between these two levels (at least one but less than 30\% of the products say it's malicious) is then discarded from the experiment \cite{Berlin:2015:MBD:2808769.2808773,Saxe2015a}. We note there is nothing particularly special about choosing 30\% as the threshold, and many works have used different rules. Others have used $\geq4$ AV hits as a splitting point between malicious and benign \cite{Incer:2018:ARM:3180445.3180449}, or left the issue unspecified \cite{Kolosnjaji2016}.

A recent study by \textcite{251586} of different thresholding strategies found that using a threshold of $\leq 15$ as the decision point between benign and malicious is a reasonable compromise to varying factors. This includes the fact that 1) AV decisions fluctuate over time, stabilizing after several months. 2) The false positive rate of AV engines in not trivial for novel files, 3) the false positive rate on packed benign files can be significantly higher, and 3) many AV engines have correlated answers and some appear to alter their own decisions based on the results of other AV engines over time. We note that these results regarding the use of VirusTotal labels are only for a benign vs malicious determination, and an equally thorough study of family labeling using VirusTotal has not yet been presented. 

While this selection is easy to perform, the labels will be intrinsically biased to what the AV products already recognize. More importantly, binaries marked by only a few AV products as malicious are likely to be the most important and challenging examples. This middle ground will consist of either benign programs which look malicious for some reason (false positives), or malicious binaries that are not easy to detect (false negatives). Removing such examples will artificially inflate the measured accuracy, as only the easiest samples are kept. Removing such difficult to label points will also prevent the model from observing the border regions of the space between benign and malicious. The aforementioned EMBER dataset uses this style of labeling, and hence the "easy" designation \cite{Anderson2018}. 
This AV-bias issue also hampers effective model evaluation, as we are skewing the data and thus the evaluation to an easier distribution of benign and malicious samples. This causes an artificially large accuracy by any of the metrics we will discuss later in \autoref{sec:evaluation}. 

Only recently have some of these AV biases been categorized and described. \textcite{Botacin2020} has shown that the detection rate of AV products may vary by country (i.e., is this malware global, or country specific, in its proliferation), executable type (e.g., COM files vs. DLL), and family type (e.g., ransomware vs trojans). These biases will naturally be embedded into any model and evaluation built from labels that are AV produced. Further, using older files to try and maximize confidence is not a guaranteed workaround, since AV engines will have label regressions over time, where they stop detecting sufficiently old files as malicious \cite{Botacin2020}. 

We also note that the subscription service to VirusTotal allows for downloading the original files based on their hash values. This is how users can get the raw version of the EMBER dataset, or create their own preliminary datasets. However, the subscription to VirusTotal is not cheap (even with academic discounts), and may be beyond the budget of smaller research groups or those just getting into this space. As such it represents an unfortunate barrier to entry, especially since VirusTotal is widely adopted within the industry. 

When the desired labels are for malware families, the use of AV outputs becomes even more problematic. 
The family
labels provided by AVs are not standardized and different AV products will often disagree on labels or type \cite{Bailey2007}. While more advanced methods exist than simple thresholding (e.g., 3/5 of AVs say the label is "Conficker") for determining benignity \cite{Kantchelian:2015:BMG:2808769.2808780} and malware family \cite{Sebastian2016}, the use of many AV products remains the only scalable method to obtain labels. High quality family labels require manual analysis, which as noted before, requires days-to-weeks of effort. Worse still, malware authors have historically copied/stolen code from one-another, which can make determining a specific family (and the related problem of attribution) even more difficult \cite{Calleja2019}.

Beyond the issue of collecting data, there is also the fact that binaries exhibit \textit{concept drift}, meaning the population as a whole changes over time. This is true of both benign and malicious binaries, as changes will percolate through the population as the Windows API changes, code generation changes with newer compilers, libraries fall in and out of favor, and other factors. It then becomes important to investigate the performance of a classification system as change occurs \cite{Masud2011}, which is not widely explored. The distribution of malware in particular drifts at a faster rate, as malware authors attempt to modify their code to avoid detection. For example, \textcite{Rajab2011} performed an extensive study of web based malware on Google's Safe Browsing infrastructure. Over four years they saw an increase in malware that relied on social engineering, 
a short lifespan for the use of most exploits documented by Common Vulnerabilities and Exposures (CVEs), and an increase in attempts at "IP cloaking" to obscure their source. The fast evolution of malware is a result of an \textit{adversarial} scenario, and only further complicates the development of a long term solution \cite{Kantchelian:2013:AAD:2517312.2517320,Singh:2012:TCD:2381896.2381910}. 

\section{Features for Binaries} \label{sec:features}

Feature extraction is the prerequisite step to applying any machine learning method. In the domain of malware analysis for PE binaries, the approaches are generally divided into one of two groups: static or dynamic. Dynamic features are extracted from running a binary and extracting the features of interest from its execution. The fact that a program's execution may alter from run to run and in different environments is why the features are called dynamic. Given the potentially malicious intent of any arbitrary executable, 
dynamic analysis is often done through a virtual machine (VM).  Conversely, static features are extracted from the binary itself without any attempt to execute it. A summary of the most commonly used features, and the representations that are regularly used, is given in \autoref{tbl:feature_summary}. 

\begin{table}[bt]
\centering
\caption{Summary of the features commonly used for malware analysis. Feature Source columns indicate whether the feature type is \textit{commonly} obtained via dynamic or static analysis. Feature Representation indicates the which ways of interpreting the original features are used. The fixed-length column does not consider cases where a approach is used that converts sequences and graphs to fixed length representations while retaining significant information of the sequential nature.  }
\label{tbl:feature_summary}
\begin{tabular}{>{\kern-\tabcolsep}lccccc<{\kern-\tabcolsep}}
\toprule
                & \multicolumn{2}{c}{Feature Source} & \multicolumn{3}{c}{Feature Representation} \\
\multicolumn{1}{c}{Feature Type}    & Static         & Dynamic         & Fixed-Length     & Sequence     & Graph    \\
                 \cmidrule(l){1-1} \cmidrule(l){2-3} \cmidrule(l){4-6}
Bytes           &  \checkmark    &                 &  \checkmark      &  \checkmark  &          \\
\rowcolor{gray!20}
Header Values   &  \checkmark    &                 &  \checkmark      &              &          \\
Entropy         &  \checkmark    &                 &                  &  \checkmark  &          \\
\rowcolor{gray!20}
Assembly        &  \checkmark    &  \checkmark     &  \checkmark      &  \checkmark  & \checkmark \\
API/Function Calls  &  \checkmark    &  \checkmark     &  \checkmark      &  \checkmark  & \checkmark \\
\rowcolor{gray!20}
System Calls    &                &  \checkmark     &  \checkmark      &  \checkmark  & \checkmark \\
Network Traffic &                &  \checkmark     &                  &  \checkmark  & \checkmark \\
\rowcolor{gray!20}
Performance Counters &                &  \checkmark     &   \checkmark     &  \checkmark  &          \\
System Changes  &                &  \checkmark     &   \checkmark     &  \checkmark  &          \\
\rowcolor{gray!20}
Contextual        &  \checkmark    &  \checkmark     &        &    &   \checkmark \\
\bottomrule
\end{tabular}
\end{table}

\subsection{Dynamic Analysis Features} \label{sec:dynamic_features}

There are a number of common feature types to extract from dynamic analysis. For example, an early type of dynamic analysis was to modify the linker in the operating system to wrap each function call to the OS or other libraries with a special prologue and epilogue \cite{Willems:2007:TAD:1262542.1262675}. In doing so the functions called could be tracked in the order of their occurrence and one could obtain a sequence of \textit{API or function calls}. Such trackings of API calls can be used in many ways, and is often interpreted as a sequential ordering or as a directed graph \cite{Elhadi2014,Fredrikson2010}.  Special tracking can be added for common tasks, such as registry edits, files created or deleted, mutex operations, and TCP/IP calls \cite{Rieck:2008:LCM:1428322.1428330}. These are all common tasks or operations that malware might perform, so recording extra information (such as method arguments) can be beneficial to analysis. Ultimately, there are many ways to combine the API functions called and the operations performed, with many works using one of or both options, and tracking different subsets of actions. These approaches are often called "behavior based", and make up a large portion of the dynamic features used. Directly related to tracking of API calls is tracking \textit{system calls}.  For our purposes, we define a system call as a service provided by the Windows kernel, and (usually) accessed via an entry point in {\tt Ntdll.dll}.  
\cite{Russinovich:2012:WIP:2464864,Russinovich:2012:WIP:2361939}
There are several hundred of these functions, and they are often called by the APIs Microsoft provides, but less often by user code.  In fact, use of functions in Ntdll.dll by user code is regarded as a malware indicator \cite{Sikorski:2012:PMA:2181153}. 
One advantage of tracking system calls, rather than all calls to the Windows API, is that the set of system calls tends to remain stable from one version of Windows to another, for the sake of compatibility.

The same technology that allows for API call traces can also be used to track changes to the state of the system. Such \textit{system changes} may include the registry edits and files created, as well as processes that started or ended, and other various configurable settings within the OS \cite{Bailey:2007:ACA:1776434.1776449,184519}. System changes may also be obtained from system logs \cite{Berlin:2015:MBD:2808769.2808773}, which can be used as a convenient feature source with minimal overhead (since the system was going to collect such logs anyway) or for retroactively detecting malware and determining the time of infection. 

Though not as popular, more granular information can be extracted as well. It is  possible to record the sequence of assembly instructions as they run \cite{Dai2009}. Though this approach in particular can require additional feature selection and processing, as the amount of data can grow quickly and the length of program execution may be unbounded. Another option is to track the results of various performance and hardware counters that are present in modern CPUs as well as process related counters tracked by the OS  \cite{Tang2014}. These could include the number of memory pages being allocated or swapped, voluntary and forced context switches, cache hits and misses, and other various fields. The intuition being that the performance behavior of malware will be distinct from benign applications due to the different nature of their operation. 

Another less frequently used approach is to monitor the network traffic and content that a binary may produce \cite{Stakhanova:2011:ENM:2358959.2359567,Nari2013,Wehner:2007:AWN:1370628.1370630,Perdisci:2010:BCH:1855711.1855737,Mohaisen:2013:UZA:2487788.2488056}. Many different malware applications make use of command-and-control servers (
the existence or location of which may be obfuscated) to direct the actions of infected hosts, making it a potentially informative behavior. 
Use of the local network is also one of the most common ways for malware to self proliferate. 
While the population of malware that does not use the Internet or any local network may be small, it may also be one of the more interesting and important ones to classify correctly. 
The methods discussed in this section
make up the majority of features that are extracted via dynamic analysis. While the set of options may seem simple, the systems to capture them represent their own significant engineering efforts. Many such systems have been developed over time, and we refer the reader to \cite{Egele:2008:SAD:2089125.2089126} for a survey of the numerous systems for dynamic analysis and their relative pros and cons. The focus of this work will remain not on the method of collection, but what is collected and the challenges that are faced in doing so.

\subsection{Dynamic Analysis Challenges} \label{sec:dynamic_challenges}

At first glance, a preference for dynamic over static features may seem obvious. The actual behavior of an application would intuitively be a strong indicator of the intent of a binary, and an effective way to group applications and measure their similarity. 
However, this perforce requires allowing the binary to execute --- which opens a proverbial can of worms that must be considered. 

For safety, malware must generally be run inside a Virtual Machine where its effects can be contained and reverted. But the malware authors are aware of this, and can attempt to detect that the malware is being run in a controlled environment and then alter the malware's behavior in response. It is even possible for malware authors to detect which specific emulator they are being run in, be it standard Virtual Machine emulation software (e.g., VirtualBox) or custom environments used by AVs \cite{Blackthorne:2016:AFA:3027019.3027028}. 
For evasive malware, that means the apparent behavior inside of a safe Virtual Machine may differ in a substantial way from when the same program is running on real hardware \cite{184519}. This makes features built from running binaries inside a VM less reliable. 

Unfortunately there exist a number of potential ways for a malicious author to detect a virtual environment, and there is no simple way to prevent such detection. One particular avenue is through CPU and timing attacks \cite{Kang2009} that are applicable to hypervisor virtualization \cite{Popek:1974:FRV:361011.361073}. For a bare-metal hypervisor that allows most instructions to run at native speed, it is necessary to intercept and emulate certain instruction calls (such as changing from ring 3 to ring 0) in order to keep the VM contained. Such instructions will incur a significant performance penalty due to the extra overhead to intercept and emulate them. While this is normally acceptable, as such cases are the minority of instructions, the performance discrepancy may be used by the binary to determine that it is running under emulation, and thus alter its behavior. Similarly, if the whole system is being emulated equally "slowly", malware could request information about the CPU, network card, and other hardware to determine if the time to execute is abnormally slow for the given hardware or inferred hardware age. Even beyond just timing attacks, the numerous possible discrepancies between real and emulated hardware have lead many to consider the task of creating a virtual-machine undetectable by malware effectively impossible \cite{Garfinkel:2007:CTV:1361397.1361403}. 

One avenue of research to circumvent this problem is to force binaries to follow some path of execution \cite{184513,Brumley2007}. Such approaches successfully avoid the issue of allowing malware to determine its own behavior, at the cost of not necessarily knowing which execution path to take to observe desired behavior. That is to say, we do not know which execution path and sequence will exhibit the malicious behavior we wish to detect. 
Even if we ignore looping constructs and backwards branches,
if a binary has $b$ conditional branches (e.g. If-Else statements) in it, there may be up to $2^b$ different possible execution paths to take. Some heuristics to select execution paths must be applied, and this may be difficult given the unusual behavior of malicious binaries. For example, one may heuristically switch execution paths if one path causes illegal behavior or results in an interrupt from the OS. However, such behavior may be intentional, in causing side effects or triggering a bug that the malware intends to exploit. 

Even given the pros and cons between execution in a virtual environment and forced execution, both approaches share a common issue in application. Behavior of malware may depend on the user environment in a non-trivial way. A trivial case would be malware behavior dependent on a bug specific to an OS version, such as Windows XP over Windows Vista. It has been found that malware may depend on specific applications being installed and running at the same time as the malware, and the interactions between programs in regular use \cite{Rossow2012}. Such scenarios are not general or easily covered in experimental testing, and can cause a large discrepancy between the lab and deployments to real users. Such cases may easily cause a machine learning model
to stop generalizing, or miss certain subsets of malware in practice.  

Another environmental factor in dynamic analysis is Internet traffic and connectivity. Allowing unconstrained Internet access to running malware is risky at best, and opens ethical concerns in allowing malware under examination to infect and attack other machines. Yet disconnecting Internet access entirely may dramatically alter the behavior of malware, not including the possibility of malware updating itself or downloading new functionality. Maximizing the containment of malware while allowing Internet access can require extensive design and engineering effort \cite{Kreibich:2011:GPC:2068816.2068854}. A further complication exists in experiment reproducibility, as the servers malware connects to may change or go offline over short periods of time. When these servers do not respond, or even if they do, the malware's behavior may change or cease altogether. This makes dynamic analysis of older malware difficult, as these servers are unlikely to return\cite{Rafique:2013:FMC:2941590.2941601}.

The issue of reproducibility and infection can be partially addressed by network emulation, in which the host environment running the malware intercepts and alters network traffic, and potentially provides fake responses, in order to let the malware run as if it had Internet connectivity while keeping it isolated \cite{Graziano:2012:TNC:2420950.2421000}. These issues are significant impediments in using network traffic as a reliable feature, and only further complicate dynamic analysis. A new approach to help make dynamic feature extraction more reproducible is to design special VM recording techniques, which save all of the non-deterministic events so that a VM can be replayed at a later point in time \cite{Severi2018}. While powerful and storage efficient, if the malware successfully detects the VM at first run and alters its behavior (or fails to run properly for other reasons), the replay will always reflect this failure.

\subsection{Static Analysis Features} \label{sec:static_features}

By its very nature, static analysis greatly reduces the scope of features options to consider for classification. One common choice is to use the raw-bytes themselves as features \cite{raff_ngram_2016,Kolter:2006:LDC:1248547.1248646,Stolfo2007}. A subset of the raw-byte approach is simply to search for and extract what appear to be ASCII strings \cite{5615149}. This approach assumes the least amount of knowledge and is widely applied to other file types because of its ease of application. Another approach is to instead compute a windowed entropy over the raw bytes, mapping each file to a entropy sequence \cite{Han2015,Baysa2013,Sorokin2011}. Regardless of how processed, these approaches have an attractive simplicity at the cost of ignoring relevant properties. For example, while the raw bytes may be processed as one long linear sequence, the locality within a binary is non-linear. Different portions will relate to others through pointers in the storage format as well as various local and long jumps in the assembly. It is also common to build histograms from this information to reduce it to a fixed length format \cite{Saxe2015a}. 

Using more domain knowledge, it is also popular to parse the PE-Header \cite{MicrosoftPEFormat} for relevant information, extracting the fields and imports and encoding them as numeric and categorical features \cite{Shafiq2009,raff2017peheader}. Being able to process the PE-Header is also important for finding and disassembling the binary code, which is one of the more popular feature types to use \cite{Santos2010,Moskovitch:2008:UMD:1485445.1485472} in static analysis. As mentioned in \autoref{sec:dynamic_features}, assembly sequences can be used in dynamic analysis as well. The difference then becomes what assembly sequences appear in the file and overall structure, versus the sequence of instructions actually run \cite{Damodaran2015}. In each case one may observe sequences not seen by the other. The dynamic version may not run all of the code present, and the static version may not find obfuscated instructions. 

The extraction of PE-Header disassembly from static analysis are more readily available, and provided by many open-source projects. For the PE-Header, there are projects like PortEx \cite{Hahn2014}, pefile\footnote{\url{https://github.com/erocarrera/pefile}}, and this functionality is even built-into the Go language runtime\footnote{\url{https://golang.org/pkg/debug/pe/}}.  For disassembly, relevant projects include Capstone\footnote{\url{http://www.capstone-engine.org/}}, Xori\footnote{\url{https://github.com/endgameinc/xori}}, Distorm\footnote{\url{https://github.com/gdabah/distorm}}, BeaEngine\footnote{\url{https://github.com/BeaEngine/beaengine}}, and others. 
Many different disassemblers have become available in part because disassembling a binary is non-trivial, especially when malware may attempt to create obscure and obfuscated byte code that attempts to thwart disassembly. Each of the many options available have different pros and cons in terms of run-time, accuracy, supported architectures, and other issues. 

Once a binary is successfully disassembled (which requires the PE Header), it is also possible to resolve API function calls from the assembly using the Import Address Table \cite{Ferrand2016}. The IAT stores the functions the library wishes to load as well as the virtual address at which the function will be stored. Then any \mintinline{nasm}{jump} or \mintinline{nasm}{call} function's arguments can be converted to the canonical target function. This allows us to not only use the imported functions and APIs as features in a fixed-length feature vector (function present / absent), but also as a sequence or graph of API call order. 

Finally, the most knowledge-intensive and time-consuming option is to consult malware analysts on what information to look for, and attempt to automatically extract said information \cite{Dube2012137}. Such approaches may obtain a distinct advantage from expert opinion, but will require additional work to update due to concept drift as malware authors adjust their code to avoid detection. 

\subsection{Static Analysis Challenges} \label{sec:static_challenges}

While the feature extraction process is often simpler for static analysis, it exhibits its own set of problems that must be dealt with. Notably the contents and intent of a binary are often obfuscated, with the first line of obfuscation being the use of packing \cite{Wei2008}. Packing wraps the original binary content inside a new version of the binary, often storing the original version with some form of compression or encryption. Upon execution, the packed version of the binary extracts the original version and then executes it. Packing may be applied recursively multiple times, with different types of packers each time, to maximize the effort needed to extract its contents. 

This technique has been widely used in malware, and among well-meaning software developers. Packing is often employed as an attempt to thwart reverse engineering by a competitor, avoid or delay the prevalence of "cracked" versions of commercial software, or just to reduce the file size for transfer \cite{Guo:2008:SPP:1433006.1433014}. There are attempts to encourage the authors of packing software to cooperate with AV vendors to add information that would reduce the magnitude of this problem \cite{6112319}, incorporating "tags" that would make it easier to determine if a packed binary is safe and where it came from. Currently it remains that it is not sufficient to simply detect packing and infer maliciousness. The development of automated unpacking tools is an active area of research \cite{Martignoni2007,Royal2006}, however it generally requires some level of emulation of the binary. This brings back many of the issues discussed in \autoref{sec:dynamic_challenges} with performing dynamic analysis. Though there has been some work in static unpacking \cite{Coogan:2009:ASU:1685994.1686221}, the dynamic approach to this problem has been the preferred method in most works. 

Packing is often considered as a "catch all" that thwarts all static analysis, and always increases the entropy of the original file. Recent work by \cite{Aghakhani2020} has challenged many of these "known" assumptions about packing. In a large study they have shown it is possible for machine learning based models to make benign vs malicious distinctions even when contents are packed, provided the packers are not too novel and the training distribution of packers is properly accounted for. They also showed that many packers lower the entropy of resulting file. A result in particular that is counter-intuitive from this work is the utility of byte $n$-grams as features. This feature type will be discussed more in \autoref{sec:features}, but many assumed packing would invalidate such byte based processing from being useful. 

There exists other types of obfuscations as well within the binary code itself, including a host of possible obfuscations done by packers \cite{Roundy:2013:BOP:2522968.2522972}. Some simpler forms of obfuscation may include the use of extra instructions that don't change the result of a program, executing instructions generated at run-time (separate from unpacking), and unnecessary or unreachable code \cite{EPFL-REPORT-167534,Christodorescu2007}. 

There also exist other sophisticated obfuscation techniques that are widely used, but from which information can be extracted with some effort. 
Polymorphic malware alters itself each time it propagates, creating numerous different versions that are all functionally equivalent while obfuscating the entry point or decryptor of a binary \cite{Newsome:2005:PAG:1058433.1059393}. Metamorphic malware goes further, potentially altering all of the binary code as it propagates \cite{RHUL-MA-2008-02}. Some malware even implements its own virtual machine for a custom instruction set in which the malicious code is written \cite{Sharif2009}. Analysis can get particularly difficult when multiple forms of obfuscation are used, since none of them are mutually exclusive. There have been attempts to develop fully generic deobfuscation systems that do not rely on knowing in advance which obfuscation technique is being used, but such attempts have not yet been fully successful \cite{Yadegari2015}. 
Granted that a competent malware analyst can reverse engineer many if not most obfuscated files, with the right tools and sufficient time, \cite{Schrittwieser:2016:PST:2911992.2886012}, such efforts are expensive and do not scale. 

It has been shown that deobfuscating malware improves the recall of signature based approaches \cite{Christodorescu2007}.  The presence of obfuscation may be a malware indicator in its own right, and such a feature could be useful in building a machine learning model.  Hence, it is not clear that deobfuscation should be attempted in each and every case, and arguments could be made either way.  This question deserves further study.

\subsection{Contextual Features}

A third type of features are what we will call \textit{contextual} features. These are features that are not properties of the malicious binary itself, but come from the context of how the malware may exist or be distributed. The use of contextual features is less common in research, but has been reported to be highly successful in practice.  Such systems are generally graph-based in their approach. 
For example \textcite{Chau2011} used information about the "reputation" of the machines at which an executable file was found to make a determination about maliciousness, without looking at the content of the file itself.  Others have followed this same strategy, and attempt to more precisely define the relations between files to improve results \cite{Tamersoy:2014:GAL:2623330.2623342}, and to merge both relations and file dependent features \cite{Ye:2011:CFC:2020408.2020448}. 

Beyond measuring reputation of machines, the reputation of the domain name or IP address from which a file was downloaded can also be used to classify the downloaded binary as malicious if the source address has low reputation. This, as well as counter-measures, were discussed by \textcite{Rajab2011}. Others have created more elaborate graphs based on how and from where the file was downloaded, including the benign applications (e.g., Internet Explorer) that are also involved in the process \cite{Kwon:2015:DEI:2810103.2813724}. 

In a similar theme, \textcite{export:193769} looked at making classifications for files that are found in the same container (e.g., a zip or rar file). This approach is based on the hypothesis that if any file found in a container is malicious, all are more likely to be malicious. A similar approach has recently been proposed to leverage the file name of the malware itself, rather than its contents, to predict maliciousness \cite{Nguyen2019_filename_malicious,Kyadige2019}. While not sufficient on its own, it may be useful in conjunction with other features \cite{Kyadige2019} or in investigative/prioritization situations where the whole file may not be available (e.g., file path was stored in a log, but the file itself has since been deleted) \cite{Nguyen2019_filename_malicious}. 

\subsection{Contextual Challenges}

The contextual information we have discussed can include a mix of both static (files rest on the same system) and dynamic sources (reputation of IP addresses and download path). As such it is not a third type of information, but the nature of the contextual information being outside of the executable makes it intrinsically different from the others we have discussed. 

The biggest impediment to using a contextual approach so far appears to be access to the contextual information itself. All of the works we have discussed make use of data owned by private companies and is measured in the millions of files, and can not be made generally available to all researchers. This makes the reproducibility and comparison issues discussed in \autoref{sec:data} especially pertinent. 

Similar to the issues discussed in \autoref{sec:dynamic_challenges}, the contextual information is sensitive to time. Unless recorded for posterity, it will not be possible to perform a historical study of how contextual features would have performed. This applies to both the static and dynamic sources of contextual information.

\section{Machine Learning Methods} \label{sec:ml_methods}

Most machine learning methods for classification work on fixed length feature vectors, i.e., $x \in \mathbb{R}^D$ where $D$ is the number of features used. This is a natural representation of information for many domains, however it is a general mismatch for the case of malware classification. With the exception of features from the PE header and some expert knowledge features, almost every feature choice discussed in \autoref{sec:features} is sequential in nature. This leaves us with two choices, both less than ideal: make some simplifications to the problem so that we obtain fixed-length feature vectors, or restrict ourselves to the more limited set of models that support classification of sequences. Below we discuss the primary method by which fixed length features are constructed, and the many algorithms that have have been used for both fixed-length vector and sequence-based classification. Other methods that more directly tackle the true nature of our feature choices will be discussed in \autoref{sec:ml_seq} and \autoref{sec:graphs}. 

A natural question to ask is which of the learning approaches and feature combinations work best. Unfortunately this question cannot be easily answered due to the data issues discussed in \autoref{sec:data}. For the case of malware detection, many results are likely overestimated, and the lack of any common benchmark dataset further hinders any attempts to compare and contrast results. When distinguishing between malware families the VX-Heaven corpus provides a shared benchmark, but it is a sub-optimal barometer. Not only is the corpus outdated to the point that it does not reflect modern malware, but no particular effort is made to balance the number of samples from each malware family. That is to say, both the types of malware and individual malware families are not evenly distributed. This makes interpretation of results more difficult, especially as many works sub-sample the corpus in different ways to remove families or malware types with fewer samples. 

Given these difficulties, for each learning algorithm we discuss the feature types and scenarios where they seem to perform well, and the situations where they perform worse or we believe their utility has been over-estimated. In addition we will give background to relevant extensions and advancements in the machine learning literature  that could be relevant to future progress in the field of malware classification, but have not yet been explored.

\subsection{N-Grams} \label{sec:ngrams}

The first item we discuss is not a learning algorithm, but a method of constructing features. N-Grams are a "bread and butter" tool for creating feature vectors from sequence information, though capture very little sequential information (and hence are included in this section). Despite this, n-grams have been widely used in malware classification, starting with the work of \cite{Abou-Assaleh2004} that connected the methods being used with those in the domain of Natural Language Processing (NLP). Since then, n-grams have been one of the most popular feature processing methods for malware classification, and have been used for processing bytes, assembly, and API calls \cite{large-scale-malware-classification-using-random-projections-and-neural-networks} into bag-of-words type models. To give a more concrete example of this process, the byte sequence \textit{0xDEADBEEF} would have the 2-grams \textit{DEAD}, \textit{ADBE}, and \textit{BEEF}. At training time all possible 2-grams would be counted, and each 2-gram found would map to an index in a high-dimensional feature vector. The feature vector for \textit{0xDEADBEEF} would have 3 non-zero values, the specific values determined by some feature weighting scheme such as TF-IDF or Okapi \cite{Robertson:1994:SEA:188490.188561}, though a binary present/absent value is popular as well.

There exists a particular desire to use larger values of $n$ for malware classification due to the limited semantic meaning contained within only 6 or so bytes or instructions. To give this context, a 6-byte-gram is not large enough to capture a whole instruction 2.4\% of the time \cite{5645851}. This is due to the variable length encoding of the x86 instruction set, and a valid x86 instruction can be up to 15 bytes in length. Similarly, an assembly 6-gram is often not sufficient to cover the behavior of a larger function. A simple function can compile to dozens of instructions, let alone more complicated functions which may easily be hundreds to thousands of instructions in length. 

While large values of $n$ are desirable, they are also computationally demanding. 
As $n$ increases, the number of possible n-grams grows exponentially. Counting and tracking these itself is expensive, and feature selection is required before deploying. As such, the use of $n > 6$ has historically been rare. Some work has been done to speedup  the collection of n-grams by approximately selecting the top-$k$ most frequent n-grams as an initial feature selection process \cite{hashgram_2018}. This is based on the observation that n-grams tend to follow a power-law distribution, and that useful predictive features tend to have a minimum frequency \cite{Luhn:1958:ACL:1662353.1662360}. Later work developed this into a probabilistic algorithm for selecting the top-$k$ $n$-grams in a faster manner with fixed memory cost, testing values of $n=8192$ \cite{Kilograms_2019}. This study found that $n \geq 64$ was surprisingly useful, and had the benefit that a malware analyst could reverse engineering the meaning of a large $n$-gram to better understand what the model had learned. Their work showed that predictive performance was maximized around $n=8$, and that $n$-gram features had a surprisingly long shelf life, still being effective in detecting benign/malicious software up to 3-years newer than the training data. 

\subsubsection{N-Gram Coalescing}

To help mitigate the computational issues with n-grams while retaining as much information as possible, approaches analogous to word stemming have been applied for both byte and assembly n-grams. In NLP stemming attempts to coalesce words with similar semantic meaning into one base form (e.g., "running", "ran", and "runner", all get mapped to "run"). This coalescing may lose important nuance, but can also benefit in a reduction to a more powerful subset of features.

For x86 assembly, the likelihood of seeing an exact match for most instructions and their operands values together is quite low. This results in extra features and may fail to match instructions that are essentially the same. The simplest workaround to this problem is to map each line of assembly to just the instruction being executed \cite{dolev2008malware,Moskovitch:2008:UMD:1485445.1485472}, e.g. \mintinline{nasm}{mov eax, 4} is mapped to just \mintinline{nasm}{mov}. 
\textcite{Shabtai2012} argued in favor of this approach, noting that the main "engine" or component of malware could be easily re-located between different versions of a file. This would change the relative offsets, and thus the operands --- causing the same code to no longer match. By removing the operands completely this issue is resolved, at the cost of specificity. It is then up to the learning method, empowered by appropriately sized assembly n-grams, to learn to detect these patterns. 

Another alternative was proposed by \textcite{Masud2008}, which balances the extremes of removing the operands of the assembly and keeping them in their entirety. They noted that an instruction will have some number of parameters and each parameter could be coalesced into a location type, either memory, register, or constant corresponding to where the value used came from: either an access to memory, directly from a register, or the immediate value from the call to an instruction. For example, the instruction \mintinline{nasm}{mov eax, 4} would be coalesced to \verb|mov.register.constant| and \mintinline{nasm}{mov [eax], 4} to \verb|mov.memory.constant|. We note that in this form it does not matter that a register was used in the first parameter, it is that the operand value came from a memory access that determined the type. 

Reducing the representation space via coalescing is intuitive and attractive, but it can also obfuscate important information depending on the task.  The instruction name itself, such as \mintinline{nasm}{cmp} is in fact already performing some level of coalescing. This is because while the assembler accepts one "\mintinline{nasm}{cmp}" instruction, this instruction will be converted to one of nine different opcodes when assembled. \textcite{Zak2017} found that "disambiguating" the specific opcode an instruction was compiled down to improved the predictive performance of assembly n-grams using both of the aforementioned forms of operand coalescing. This however was only for static analysis, and results may differ when instructions are extracted in a dynamic manner.  

For bytes and assembly, n-perms have been proposed as an alternative scheme \cite{Karim2005,walenstein2007exploiting}, particularly for clustering malware. An n-perm represents an n-gram and all possible permutations of an n-gram. An equivalent re-phrasing is: to map every n-gram to a canonical n-perm based on the contents of the n-gram, ignoring their order (e.g., \textit{ACB} and \textit{BCA} would both map to \textit{ABC}). This conflation dramatically reduces the number of features created as $n$ increases, allowing the consideration of larger values of $n$. This same notion has been re-developed for assembly as well \cite{Dai2009}, as a way to circumvent metamorphic malware which will re-order instructions and add superfluous instructions as a form of obfuscation. 

\subsection{Linear Models} \label{sec:linear}

One of the simplest and most effective classes of machine learning methods are linear models, the general objective function for which is given in \eqref{eq:linear_model}. In it we have $N$ data-points, a label $y_i$ for each data-point, and a weight vector $w$ that defines the solution. The value that $w$ takes is dependent on the loss function $\ell$, the regularizing function $R(w)$ and its weight $\lambda$. The basic goal is to assign every feature $D_j$ an importance weight $w_j$, that is positive or negative depending on which class it is an indicator of. Given a positive and negative class (malicious and benign), we obtain a classification decision  by examining the sign of the dot product $\sign(w^T x)$, which is between the weight vector $w$ and a data point $x$. Despite their simplicity, linear models can be highly effective, Especially when dealing with high dimensional data sets, where more sophisticated non-linear models may have only minimal improvement in accuracy \cite{Chang2010,DBLP:journals/pieee/YuanHL12}. 

\begin{equation} \label{eq:linear_model}  
\sum_{i=1}^N \ell( w^T x_i, y_i) + \lambda R(w)
\end{equation}

For the loss function $\ell$ the two most common choices are the Logistic loss \eqref{eq:loss_log} and the Hinge loss \eqref{eq:loss_hinge}. The Logistic loss corresponds to performing Logistic Regression, and the Hinge loss corresponds to using a Support Vector Machine (SVM) \cite{Cortes1995}. As presented below the value $y$ indicates the true label for a data-point, and the value $s$ would be the raw score for the data-point --- the dot product between the weight vector $w$ and the feature vector $x$. 

\begin{subnumcases}{\ell(s, y) =} \label{eq:loss_log}
                    \log(1 + \exp(-y \cdot s))  & \text{Logistic}\\
                    \max(0, 1-y \cdot s) & \text{Hinge}
\label{eq:loss_hinge}
\end{subnumcases}

When training a linear model, the choice of $\ell$ does not have a significant impact on accuracy or training time. The choice of the regularizing function $R(w)$, and the amount of regularization we apply, $\lambda$, have a much more impact on model performance. For $R(w)$ the $L_2$ norm ($R(w) = \frac{1}{2}\norm{w}^2_2$)\footnote{The $\frac{1}{2}$ term is included because it makes the math slightly more convenient when deriving the update. Otherwise it is of no significance, and is sometimes rolled into the value of $\lambda$} 
is the most common, 
and a search over penalties $\lambda$ is done to find the value that best prevents overfitting to the training data. By increasing the value of $\lambda$ we increase the penalty for model complexity, and encourage $w$ to approach $\vec{0}$. 

The other common choice of regularizer, the $L_1$ norm ($R(w) = \norm{w}_1$), is also a potentially useful choice, 
especially when dealing with high dimensional data that can result from the use of n-grams. This is often called Lasso regularization \cite{Tibshirani1994} and will result in exact zeros occurring in the weight vector $w$, meaning it performs its own feature selection as part of training. When a hard zero is assigned as a coefficient, the associated feature has no possible impact on the model, and can be removed. 

Lasso regularization also comes with theoretical and practical robustness to extraneous and noisy features \cite{Ng2004}, where a model trained with $L_1$ regularization will perform better than one trained with $L_2$ regularization as more and more unrelated features are added. This makes it an excellent fit for n-gram based feature vectors, which can quickly reach $D >$ 1 million and has been successfully applied to byte n-grams to improve accuracy and interpretability \cite{raff_ngram_2016}.  The $L_1$ norm does have some weaknesses: it's a biased estimator and can reduce accuracy under certain situations. But $L_1$ can be combined with the $L_2$ norm to form what is known as Elastic-Net regularization ($R(w) = \frac{1}{2} \norm{w}_1 + \frac{1}{4} \norm{w}^2_2$) \cite{Zou2005}. The Elastic-Net often provides the best of both worlds, resulting in models with higher accuracy and retaining the sparsity benefits of Lasso. 

The simplicity of linear models provides the practical benefit of many tools being publicly available and able to scale to large datasets. The popular LIBLINEAR library supports both $L_1$ and $L_2$ regularized models for both the Hinge and Logistic loss functions \cite{Fan2008}. LIBLINEAR uses exact solvers specially designed for each combination of loss $\ell$ and regularizer $R(w)$. Similarly the library Vowpal Wabbit \cite{Langford2007} implements the same algorithms using online methods, meaning it trains one datapoint at a time and can stream the data from disk. This allows Vowpal Wabbit to train faster and scale to terabyte size corpora. While the online training approach may sometimes result in lower accuracy than the approaches used in LIBLINEAR, the difference is usually minimal (if there is a noticeable difference at all). Linear models are also attractive because they are fast to apply, making them realistic for the real-time goals of AV systems.

\subsection{Kernel Methods} \label{sec:kernel_methods}

Kernel methods are an extension of the linear methods discussed in \autoref{sec:linear}. Most commonly used with Support Vector Machines (SVMs) using a kernel trick $K$, the objective function is given in \eqref{eq:svm_kernel}.  We note that for SVMs the regularization penalty is usually expressed with the term $C$, where larger values indicate less regularization. These forms are equivalent where $C =  1 / (2 \lambda N)$. The kernel-trick is used to project the original data-set into a different space. A linear solution is found in this new space, which may be non-linear in the original space. 

\begin{equation} \label{eq:svm_kernel}  
\sum_{i=1}^N \max( 0, 1- y_i K(w, x_i)) + \lambda \norm{w}^2_2
\end{equation}

A valid kernel $K$ represents the inner product in this new feature space, but does not require us to explicitly form it\footnote{The kernel trick is usually more formally explained as a Reproducing kernel Hilbert space (RKHS). We avoid it to reduce the mathematical background needed for this review}. This allows us to obtain classifiers that are non-linear in the original feature space (and thus potentially achieve a higher accuracy). We can always pick the linear kernel \eqref{eq:linear_kernel}, which results in a linear model. Practically, two of the more common choices are the polynomial \eqref{eq:poly_kernel} and Radial Basis Function (RBF) \eqref{eq:rbf_kernel} kernels. The polynomial kernel is particularly helpful to illustrate the intuition behind the kernel-trick, as we can easily compute $(\alpha + \beta)^{10}$ with two operations, an addition and an exponentiation. This is computing the inner product in the polynomial space explicitly, but avoids actually expanding the polynomial. If were were to explicitly form the feature space first by expanding the polynomial, we would end up performing 36 exponentiations, 10 additions, and 38 multiplications. 

\begin{subnumcases}{K(a, b) =} \label{eq:linear_kernel}
                    a^T b  & \text{Linear}\\
                    \label{eq:poly_kernel}
                    (a^T b + c)^p & \text{Polynomial} \\
                    \label{eq:rbf_kernel}
                    \exp\left(-\gamma \norm{a-b}^2\right) & \text{RBF}
\end{subnumcases}

The price for this flexibility is generally computational, as solving the kernelized version can take $O(N^3)$ time and $O(N^2)$ memory. On top of that, a parameter search must be done for the values (such as $\gamma$) used in the kernel. This is in addition to the regularization penalty $C$. Most malware data-sets being used are on the order of 40,000 samples or less, which is still in the range of available tools like LIBSVM \cite{Chang2011}. More advanced techniques that do not attempt to obtain the exact solution also exist, allowing the use of kernel methods to larger data-sets \cite{Engel2004,Hsieh2014}. 

One of the challenges with the malware domain is the multiplicity of feature options and potential representations. For most machine learning techniques it is necessary to reduce these down to a single feature vector of fixed length for each data-point. This often results in an over-simplification for the malware domain. The use of more sophisticated kernels to alleviate this problem is a yet unexplored possibility. For example, one challenge is that the PE format specifies many different section types, the most common being sections for imports, exports, binary code, and data. However any of these section types may occur in a binary with any multiplicity\footnote{up-to a hard limit on the number of sections specified by the PE format} (e.g., one could have five different executable sections). The standard approach, if differentiating between sections, is to operate as if all instances of a section type were a part of one section. Instead, one could use a kernel that matches sets of feature vectors \cite{Grauman2005,NIPS2009_3874}, allowing the model to learn from these directly. 

Kernels can also be defined directly over strings \cite{Lodhi2002,Leslie2002}, which could be useful for comparing the function names defined within an executable or in handling unusual data content, such as URLs that can be found within malware \cite{raff_ngram_2016}. To handle the function graphs that may be generated from dynamic analysis, kernels over graphs may also be defined \cite{Vishwanathan:2010:GK:1756006.1859891,Neuhaus:2007:BGG:1543687} and has seem some small amount of use for malware classification \cite{Anderson2011}. Furthermore, the composition of multiple kernels via additions and multiplications also forms a new and valid kernel. This would provide a direct method to incorporate multiple modalities of information into one classifier. For example, we could combine a kernel over graphs on API call sequences, a linear kernel for assembly n-gram features, and a kernel over strings found in the file into one larger kernel. However these options with kernels are largely unexplored for the malware classification problem. 

\subsection{Decision Trees}

Methods based on Decision Trees have been popular in machine learning, and a number of variants and extensions to them exist. The two most popular base decision tree algorithms are C4.5 \cite{Quinlan1993} and CART \cite{Breiman1984}. A number of desirable properties have resulted in their widespread use among many domains, making them some of the most widely used algorithms in general \cite{Wu2007}. In particular, decision trees are able to handle both categorical and numeric features simultaneously, are invariant to shifts or re-scaling of numeric features, can handle missing values at training and test time, are fast to apply at test time, and often obtain competitive accuracies while requiring minimal parameter tuning. 

All of these properties can be highly relevant to malware classification, where a mix of numerical and categorical features may be common, and there are often real-time requirements for deployment on user machines. Missing values can be a common issue as well, as obfuscations performed by the malware may prevent successful extraction of a given feature. For these reasons many researchers have used decision tree based methods for malware classification \cite{Perdisci2008,Dube2012137,Anderson2018}. They are easy to apply, provided as a standard tool in most machine learning libraries across several languages \cite{scikit-learn,Hall2009,gashler_waffles,JMLR:v17:15-237,JMLR:v18:16-131,Bifet2010} and with many stand-alone tools dedicated to more powerful extensions \cite{xgboost,Wright2015,NIPS2017_6907}. 

\begin{figure}[ht]
\centering
\adjustbox{max width=\linewidth}{

\forestset{
    .style={
        for tree={
            base=bottom,
            child anchor=north,
            align=center,
            s sep+=1cm,
    straight edge/.style={
        edge path={\noexpand\path[\forestoption{edge},thick,-{Latex}] 
        (!u.parent anchor) -- (.child anchor);}
    },
    if n children={0}
        {tier=word, draw, thick, rectangle}
        {draw, rectangle, thick, aspect=2},
    if n=1{%
        edge path={\noexpand\path[\forestoption{edge},thick,-{Latex}] 
        (!u.parent anchor) -| (.child anchor) node[pos=.2, above] {Y};}
        }{
        edge path={\noexpand\path[\forestoption{edge},thick,-{Latex}] 
        (!u.parent anchor) -| (.child anchor) node[pos=.2, above] {N};}
        }
        }
    }
}

\begin{forest} 
[Certificate table's size $\geq 2 \cdot 10^4$ bytes
    [File size $\geq 10^6$
        [Malware] 
        [Benign] 
    ]   
    [is DLL?
        [is 64bit?
            [Benign] 
            [Malware] 
        ]   
        [Malware] 
    ]   
] 
\end{forest}
}
\caption{A hypothetical decision tree. }
\label{fig:decision_tree_example}
\end{figure}
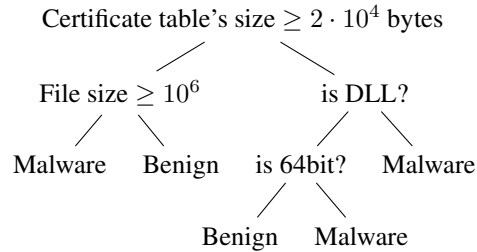

\textcite{Kolter:2006:LDC:1248547.1248646} used boosted decision trees in their seminal byte n-gramming paper. Boosting is one of many ensemble methods that work to improve the accuracy of decision trees by intelligently creating a collection of multiple trees, where each tree specializes to a different subset of the data. While they chose the AdaBoost algorithm because it performed best on their data, \author{Kolter:2006:LDC:1248547.1248646} were able to utilize the interpretability of decision trees to gain insights to their model. An example of how one would be able to read a decision tree is given in \autoref{fig:decision_tree_example}.

\textcite{raff2017peheader} used the Random Forests \cite{Breiman2001} and Extra Random Trees \cite{Geurts2006} ensembles to naturally handle the many disparate value types found within the PE-Header. Values from the header can be binary variables, multi-label variables, and numeric values with varying sizes and scales. For example, some values in the header will give the byte offset to another part of the binary, which could be anywhere from a few hundred to millions of bytes away. Most algorithms would have difficulty learning from this value range, and it can be difficult to normalize effectively. \author{raff2017peheader} also exploited the tree based approaches to obtain ranked feature importance scores \cite{breiman2003manual,NIPS2013_4928}, another method by which one can glean information about what a decision tree has learned. 

Some have worked on developing enhanced ensembling methods for decision trees to improve malware classification accuracy \cite{Menahem:2009:IMD:1497631.1497723}. Even when there is no reason to use a tree based approach in particular, many malware classification works still include them as one of many models to try \cite{Elovici:2007:AML:1421450.1421456,Alshahwan2015,Masud2008,Menahem:2009:IMD:1497631.1497723,Moskovitch2009a,Anderson2018}. 

The widespread success of decision trees has made them a valuable tool inside and outside the domain of malware classification. This has lead to a large literature of decision tree techniques to tackle various problems, many of which may be applicable to malware classification. For example, the popular AdaBoost algorithm often overfits in the presence of significant labeling errors. An extension known as Modest AdaBoost is more robust to this issue, and may lead to improvements \cite{Vezhnevets2005} in generalization. Another possibility is to use decision trees to deal with concept drift. While malware datasets with date-first-seen labels are not publicly available\footnote{Such information can be obtained from the paid-for API of VirusTotal}, there already exists a literature of decision tree methods designed to work with changing data streams \cite{Hulten2001}. This also relates to how the accuracy of a malware classification system should be evaluated, which we will discuss further in \autoref{sec:evaluation}. 

\subsection{Neural Networks} \label{sec:neural_nets}

Neural Networks have seen a recent resurgence in the machine learning community. Though older literature often referred to the technique for classification as Multi-Layer Perceptrons, newer work has placed an emphasis on the depth of trained networks and is often referred to as Deep Learning. We will provide a brief overview of neural networks, and refer the reader to \textcite{Goodfellow-et-al-2016-Book} for a more thorough introduction to modern architectures, activations, and training methods. 

\begin{figure}[ht]
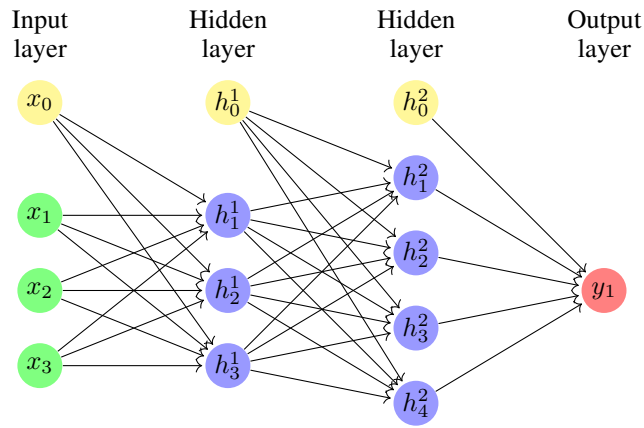
 
\centering
\begin{neuralnetwork}[height=5,toprow=true]
		\newcommand{\nodetextclear}[2]{}
		\newcommand{\nodetextx}[2]{$x_#2$}
        \newcommand{\nodetextha}[2]{$h^1_#2$}
        \newcommand{\nodetexthb}[2]{$h^2_#2$}
		\newcommand{\nodetexty}[2]{$y_#2$}
		\inputlayer[count=3, bias=true, biaspos=top row, title=Input\\layer, text=\nodetextx]
		\hiddenlayer[count=3, bias=true, biaspos=top row, title=Hidden\\layer, text=\nodetextha] \linklayers
        \hiddenlayer[count=4, bias=true, biaspos=top row, title=Hidden\\layer, text=\nodetexthb] \linklayers
		\outputlayer[count=1, bias=false,biaspos=top row, title=Output\\layer, text=\nodetexty] \linklayers
	\end{neuralnetwork}
    \caption{Diagram of a simple 1-layer neural network. {\color{green} Green} nodes are input features. {\color{yellow}Yellow} nodes are for the bias variable. {\color{blue} Blue} nodes are hidden layers. {\color{red} Red} nodes are the output layer.}
    \label{fig:fully_connected_nn}
\end{figure}

Neural networks get their name from their original inspiration in mimicking the connections of neurons in the brain (though the interpretation is often taken too literally). A neuron is connected to multiple real valued inputs by a set of synaptic weights, corresponding to real-valued multipliers. An activation function $f(x)$ is used to produce the output of the neuron, where the input is the weighted sum of every input connected to that neuron. Generally the initial features fed into a network are called the input layer, and a set of neurons that produce our desired outputs form the output layer, from which a loss is derived (such as cross-entropy, or mean squared error). In-between is some number of hidden layers, where we specify the number of neurons, connectivity, and activations for each layer. The classic approach is to connect every neuron in one layer to every neuron in the preceding layer to form a fully connected network. A diagram of this arrangement is presented in \autoref{fig:fully_connected_nn}. 

The weights for such a network are learned through an algorithm known as back-propagation \cite{Rumelhart1986}, which is performing gradient decent on the function created by the neuron graph. The view of neural networks as a large function graph has become increasingly popular, and allows for fast development using Automatic Differentiation. The user specifies the functions used by the network, and the software automatically computes the gradients with respect to any weight in the network. This has helped to fuel the resurging neural network literature and is a feature supported by many deep learning frameworks \cite{chainer_learningsys2015,chollet2015keras,Abadi2016}. 

The fundamental strategy enabled by such an approach is that the user should avoid feature engineering, and instead alter the network architecture to the needs of the problem. This works as neural networks have been found to learn their own complex feature hierarchies and representations from raw data alone \cite{ICML2012Le_73,goodfellow2014generative}. This ability has allowed neural networks to become the state of the art in both speech processing~\cite{graves2006connectionist} and image classification~\cite{NIPS2012_4824}, significantly outperforming prior approaches. 

While many works in malware classification have made use of neural networks \cite{5675808,Perdisci2008,6566472,Hardy2016}, they are often based on dated approaches to neural network construction. Advances in optimization algorithms (gradient descent), activation functions,  architecture design, and regularization have dramatically changed the "best practices" of neural networks while also improving their performance. 

One of the first effective applications of a modern neural network style was by \textcite{Saxe2015a}, who used a private corpus to obtain reasonable accuracies.  Their work performed the feature engineering manually by processing a combination  of entropy, string, and PE header features. While their model performed well, it was not compared with any other machine learning algorithms, making it difficult to determine the benefits of neural networks in this particular application. The work of \textcite{Saxe2015a} is also notable for its model evaluation, which we will discuss further in \autoref{sec:evaluation}. 

\textcite{raff2017peheader} also looked at using a neural network, but instead presented it with raw byte information and did no feature engineering. Their work provided initial evidence that neural networks can reproduce the same feature learning on raw bytes, but was limited to a relatively small (and fixed size) subset of the PE header. 
As \textcite{raff2017peheader} noted, the behavior and locality within malware is markedly different from signal and image processing tasks. Malware lacks the rigid scope and especially locality properties these other fields enjoy. As an example, it is easy to see how in an image the correlation between a pixel and its neighbors is relatively consistent throughout any image. But for a binary jumps and function calls can directly connect disparate regions, causing correlations between potentially arbitrary locations. No work has yet been done on determine what kinds of architectures can learn best from this type of locality complexity. 

Another interesting application of modern neural network design is by \textcite{Huang:2016:MMN:2976956.2976984}. Their system looked at System-call like features extracted via dynamic analysis, reducing the feature set by applying domain knowledge about the relationship function calls. Their architecture was modified to perform both malware detection (benign or malicious) and family classification (with 100 malware families) simultaneously. The jointly trained model resulted in a relative improvement of 26\% over a model trained to do only malware detection on the same data. This shows the potential for intelligent architecture design choices to provide real gains in accuracy. This is also a method to enable more data use for training, as it is easier to get more malware data labeled with malware family labels than it is to get more benign data. The portion of the network trained to predict malware family can then be trained with this additional data, without biasing the malware detection portion of the network due to class imbalance.

\section{Machine Learning Over Sequences} \label{sec:ml_seq}

Many of the feature types we have discussed, such as assembly instructions, can be more accurately described as a sequence of events or values. Using n-grams to convert them to fixed length feature vectors allows us to use the wider breadth of models discussed in \autoref{sec:ml_methods}, at the cost of ignoring most of the sequential nature intrinsic to the data. 
In this section we will review a number of techniques that are designed specifically for processing sequences, some of which will work directly on the sequence level, while others may attempt to create fixed-length representations more intelligently. In the latter case, the primary difference compared to the n-gram approaches discussed in \autoref{sec:ngrams} is that n-grams only capture a very small amount of local sequence information. Approaches in this section will more generally capture larger amounts of the sequential structure in the data. 

Some of the methods we will talk about face unique challenges regarding sequence length. For example, assembly sequences from binaries can be hundreds of thousands of steps in length or more, which significantly outpaces the longest sequences in other domains. While byte and assembly sequences are obviously the longest, potentially millions of steps long, higher level events and features extracted via dynamic analysis can easy reach hundreds of thousands of steps in length \cite{export:249072}. These sequences are far longer than what machine learning is normally applied to, meaning the tools to tackle this problems are often lacking. For example, the longest sequence length we are aware of for neural networks outside of malware is in audio generation. Here the WaveNet architecture was applied to a sequence length of 16,000 steps \cite{wavenet}. This was an order of magnitude larger than what others had achieved, yet is still an order of magnitude smaller than the sequences in the malware space. 

\subsection{Hidden Markov Models} \label{sec:hmm}

Given the sequential nature of our feature choices, such as byte sequences, instructions, and API calls, the Hidden Markov Model (HMM) has become a popular choice \cite{Damodaran2015,Shafiq2008,Wong2006,RHUL-MA-2008-02}, as it explictly models sequences and can handle sequences of variable lengths. 
Given a sequence of observations $\mathcal{O} = O_1, O_2, \ldots, O_T$ that are discrete ($O_t \in \{1, \ldots, K_O\}$), HMMs make three simplifying assumptions: 
\begin{enumerate}
\item A sequence of hidden (or unobserved) variables $\mathcal{X} = X_1, X_2, \ldots, X_T$, where each state is discrete ($X_t \in \{1, \ldots, K_X\}$) governs the observed values $\mathcal{O}$.
\item Each observation $O_t$ is generated solely by state $X_t$, and $P(O_t | X_t)$ is parameterized by a $K_X \times K_O$ emission matrix $B$. 
\item Each hidden state $X_t$ is generated solely by state $X_{t-1}$, and $P(X_t | X_{t-1})$ is parameterized by a $K_X \times K_X$ emission matrix $A$. The first hidden state is a special case, and is specified by an initial probability matrix $\pi$.
\end{enumerate}

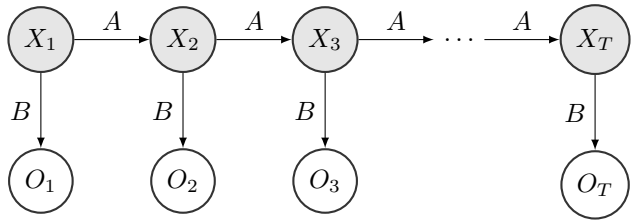
\begin{figure}[ht]
\centering
\begin{tikzpicture}
  \tikzset{
    main/.style={circle, minimum size = 5mm, thick, draw =black!80, node distance = 10mm},
    connect/.style={-latex, thick},
    box/.style={rectangle, draw=black!100}
  }
  \node[main,fill=black!10] (L1) [] {$X_1$};
  \node[main,fill=black!10] (L2) [right=of L1] {$X_2$};
  \node[main,fill=black!10] (L3) [right=of L2] {$X_3$};
  \node[main,rectangle,draw = white] (Ld) [right=of L3] {$\ldots$};
  \node[main,fill=black!10] (Lt) [right=of Ld] {$X_T$};
  \node[main] (O1) [below=of L1] {$O_1$};
  \node[main] (O2) [below=of L2] {$O_2$};
  \node[main] (O3) [below=of L3] {$O_3$};
  \node[main] (Ot) [below=of Lt] {$O_T$};
  \draw[-latex] (L1) -- node[above] {$A$} (L2);
  \draw[-latex] (L2) -- node[above] {$A$} (L3);
  \draw[-latex] (L3) -- node[above] {$A$} (Ld);
  \draw[-latex] (Ld) -- node[above] {$A$} (Lt);
  \draw[-latex] (L1) -- node[left] {$B$} (O1);
  \draw[-latex] (L2) -- node[left] {$B$} (O2);
  \draw[-latex] (L3) -- node[left] {$B$} (O3);
  \draw[-latex] (Lt) -- node[left] {$B$} (Ot);
\end{tikzpicture}
\caption{First-Order Hidden Markov Model, hidden states are gray and observed states are white.}
\label{fig:hmm}
\end{figure}

Thus the matrices $A, B, $ and $\pi$ fully specify a HMM, an example of which is given in \autoref{fig:hmm}. $A$ and $B$ are emission matrices, where the row index $r$ corresponds to the source token being $r$, and the column $c$ indicate the probability of token $c$ preceding token $r$. Because these are generative models, to apply HMMs for malware classification a separate HMM must be fit to each class from the sequences corresponding to that class. 
At test time we get a new sequence $\tilde{\mathcal{O}}$, and compute the probability $P(\tilde{\mathcal{O}} | A_i, B_i, \pi_i)$ for each HMM we learned, choosing the class with the highest probability\footnote{As presented this assumes each class is equally likely. This is generally not the case}.  
For a more thorough review of HMMs we refer the reader to \cite{Rabiner:1990:THM:108235.108253,ghahramani2001introduction}.

One can construct $m$'th-order HMMs to try to capture more history. However the learning process scales at $O({K_O}^{m+1})$, which quickly becomes computationally intractable. This makes HMMs a better fit to shorter sequences such as API call sequences, if they are to be used. It may also be the case that the use of HMMs, as generative models, make the learning problem harder than necessary. Given data $x$ and labels $y$, it models the joint distribution $P(x, y)$, which may be more complicated than the posterior distribution (i.e., benign or malicious) $P(y | x)$. While it is not always the case that discriminative models perform better \cite{NIPS2001_2020}, we suspect that it is likely true for malware classification given that just modeling a binary (i.e., $P(x)$) is its own highly complicated problem, and the joint distribution is intrinsically harder still. 

Deploying a solution using HMMs over bytes or instructions could also be problematic when we consider that malware evolves in an adversarial manner. It is  not difficult for polymorphic code to generate assembly matching a specific low order transition distribution \cite{Song2010}, which would allow a malicious author to generate binaries matching a particular distribution. 

\subsection{Byte Similarity Measures} \label{sec:byte_sim}

A number of methods have been used that seek to measure the similarity between two arbitrary byte sequences. These methods make no assumptions about the contents of formating of the underlying bytes, and thus can be used for any arbitrary byte input. This makes them attractive for malware classification, as they can simply use the raw binary as the target for similarity measures. Classification is then done by doing a nearest neighbor search for the most similar known binaries, and assigning a benign/malicious label based on the $k$ nearest neighbors' labels. We note here three methods that have been used for malware analysis and detection to varying degrees.

\subsubsection{Similarity Digests}

One method readily available for such work is the use of \textit{similarity digests} from the domain of digital forensics\cite{Harichandran2016}. These digests are similar to an MD5 checksum, but seek to minimize the change in output hash for changes in the input sequence. These techniques provide a similarity measure between the hash functions, which can then be used as the similarity between the digests. The state-of-the-art in this domain is the sdhash algorithm\cite{2009:BBS:1488734.1490103,Roussev2010}. The faster ssdeep is also popular, but is not as accurate as sdhash for most problems\cite{Kornblum2006}. Ssdeep in particular is unlikely to be useful for malware analysis, as ssdeep is sensitive to content-reordering. Since PE files and binary programs make regular use of pointers and can be almost arbitrarily reordered, this is a significant problem. Although sdhash does not suffer from this same limitation, its use for malware detection has been only moderate. 

While these digests have been highly successfully in their original domain, searching for similar files in a known database, they have not performed as well for malware classification. Some have found that their results can be improved by changing how the similarity scores are computed\cite{191669}, in order to eliminate undesirable properties. For example, the original scoring method did not guarantee symmetry (i.e., $\text{sim}(a, b) \neq \text{sim}(b, a)$). However, they have been useful in related problems to malware detection, such as finding code-reuse among malware\cite{Upchurch2016}. 

\subsubsection{Normalized Compression Distance} \label{sec:ncd}

The Normalized Compression Distance (NCD)  \cite{Li2004} is a distance function based on the ability of a compression algorithm to compress two sequences to their smallest possible size. The NCD distance is defined in \eqref{eq:ncd}, where $C$ is a function that returns the compressed length of an input, and $ab$ represents the joining of sequences $a$ and $b$ (normally done by simple concatenation). Given an oracle that can perform perfect compression, the NCD is a valid distance metric. This oracle cannot exist, and so NCD must be approximated using algorithms such as LZMA. Because NCD works based on compression, it can be applied to a wide variety of features and has become quite popular, especially for discovering malware families and relationships. It has been used successfully on raw bytes \cite{Wehner:2007:AWN:1370628.1370630} and on behavior sequences from dynamic analysis \cite{Bailey2007}. 

\begin{equation}\label{eq:ncd}
\text{NCD}(a,b) = \frac{C\left(ab\right)-\min\left(C(a), C(b)\right)}{\max\left(C(a), C(b)\right)}
\end{equation}

The flexibility of NCD, in that it can be applied to anything encodable as a sequence of bytes, makes it a powerful tool given the multiple different features we may wish to extract. NCD has also seen considerable use for malware detection due to its accuracy, which improves with the quality of the compression algorithm used. Yet the limits of existing compression algorithms also mean that NCD has difficulty in the face of long sequences \cite{cebrian2005common}, causing the distance metric to break down. Attempts to improve NCD have been made by changing how the concatenation $ab$ 
is done in practice \cite{Borbely2015}, but room for improvement still exists. When sequences are short enough that NCD works well, a yet unexplored possibility is to use it with some of the kernels discussed in \autoref{sec:kernel_methods}. A simple merging would be to replace the Euclidean distance in the RBF kernel with the result of NCD, producing $K(a, b) = \exp\left(-\gamma \cdot \text{NCD}(a, b)^2\right)$. 

\subsubsection{Lempel-Ziv Jaccard Distance} \label{sec:lzjd}

Inspired by the successes and background of NCD, the Lempel-Ziv Jaccard Distance (LZJD) has been developed for malware classification \cite{raff_lzjd_2017}. It draws from the observation that the compression algorithms that tend to work best in this domain make use of the Lempel-Ziv dictionary encoding scheme\cite{Alshahwan2015,Borbely2015}, and that the compressed output itself is never used. Instead LZJD creates the compression dictionary, and then measures the similarity of binaries using the Jaccard distance \eqref{eq:jaccard_sim} between the dictionaries. 

\begin{equation}\label{eq:jaccard_sim}
J(A, B) = \frac{|A \cap B |}{|A \cup B|}
\end{equation}

This alone was shown to be more accurate for nearest-neighbor classification of malware, and can be made nearly three orders of magnitude faster than NCD through the use of min-hashing\cite{Broder:1997:RCD:829502.830043}, thus alleviating the runtime cost of NCD. In addition to being faster, LZJD retains the distance metric properties lost by NCD\cite{raff_lzjd_2017}. The use of the Jaccard similarity / distance also means that it is a valid kernel, and can be directly  applied to the methods discussed in \autoref{sec:kernel_methods}. 

\textcite{raff_shwel} developed a method of converting LZJD's dictionary into a fixed length feature vector using a technique known as the "hashing trick" \cite{Weinberger2009a,Li:2012:OPH:2999325.2999482}. More interesting is their observation that the compression dictionary is sensitive to byte ordering, and single byte changes can cause large changes to the dictionary. They exploited this weakness to develop an over-sampling technique for tackling class imbalance, an issue we will discuss further in \autoref{sec:class_imbalance}. This was later refined to require less hyper-parameter tuning for easier use \cite{pylzjd-proc-scipy-2019}.

LZJD represents an approach of applying the intuition of NCD to a specific compression algorithm, the Lempel Ziv approach. Another approach along this theme is the Burrows Wheeler Markov Distance (BWMD), which again applies the intuition of NCD to the Burrows Wheeler compression algorithm \cite{Raff2020}. The Burrows Wheeler method is not as effective a compression approach as Lempel Ziv, and is reflected in BWMD not being quite as accurate as LZJD in nearest neighbor search. The benefit from BWMD comes from it producing a euclidean feature vector, rather than a a digest like LZJD does. This makes BWMD compatible with a wider class of ML algorithms, which showed how BWMD could produce better clustering and orders of magnitude faster search by leveraging more appropriate algorithms that require euclidean feature vectors \cite{Raff2020}.

\subsection{Convolutional and Recurrent Neural Networks} \label{sec:rnn}

As we discussed in \autoref{sec:neural_nets}, neural networks have become popular algorithms and can be viewed as a graph defining a large and potentially complicated function. This flexibility allows them to be extended to sequence tasks by replicating the same network for every step in the sequence. This is often referred to as "weight sharing", and leads to the idea of the Convolution Neural Network (CNN)\cite{LeCun:1989:BAH:1351079.1351090}. 
The success of CNNs in both image and signal processing has been long recognized \cite{LeCun:1998:CNI:303568.303704}. 
CNNs embed a strong prior into the network architecture that exploits the temporal/spatial locality of these problems. The convolution operator essentially learns a neuron with a limited receptive field, and re-uses this neuron in multiple locations. Thus a neuron that learns to detect edges can be reused for each part of the image, since edges can occur just about anywhere in an image. This property is not a perfect fit for malware sequence problems, and it remains to be seen if they will be useful despite the structural mismatch. CNNs may be a better fit at higher levels of abstraction, and have been applied to system call traces \cite{Kolosnjaji2016}. We also note that on their own, convolutions do not completely handle the variable length problem that comes with the malware domain. 

One common method of dealing with variable length sequences is to further extend the weight sharing idea, by adding a set of connections from one time step to the next, using the previous timestep's activations as a summary of everything previously seen. This high-level idea gives rise to Recurrent Neural Networks (RNNs), and we refer the reader to \textcite{Lipton2015} for a deeper introduction to the history and use of RNNs. We note that the CNN and RNN encode different priors about the nature of sequences, and can be used together in the same large architecture, or be kept disjoint. We will again refer the reader to \textcite{Goodfellow-et-al-2016-Book} for a broader background on neural networks. Below we will give only 
high-level details pertinent to models used in malware classification literature. 

Naive construction of a RNN often leads to difficulties with both vanishing and exploding gradients \cite{Bengio:1994:LLD:2325857.2328340}, making the training process difficult. One older solution to this problem is the Echo State Network (ESN) \cite{Jaeger2001a}. The ESN circumvents the RNN learning difficulty by selecting the recurrent weights via a stochastic process, so that no learning of the recurrent portion is needed. This may also be interpreted as a stochastic process by which we convert varying length sequences to fixed length feature vectors, after which any of the techniques discussed in \autoref{sec:ml_methods} may be applied. The parameters that control the stochastic process can be adjusted to sample different types of ESNs, and cross validation can be used to select the hyper-parameters that worked best. This simple strategy has worked well for many problems, and can be applied to a number of different types of learning scenarios \cite{Lukosevicius2012}. The ESN has been used by \textcite{export:249072} to process a set of high-level events, including API calls, for malware classification and found the ESNs to have an accuracy rate almost twice as high as an n-gram based approach. 

\tikzstyle{decision} = [diamond, draw, text width=4.5em, text badly centered, node distance=3cm, inner sep=0pt]
\tikzstyle{input} = [rectangle, text width=3.5em, text centered, rounded corners, minimum height=2em]
\tikzstyle{hidden_layer} = [circle, draw, text centered, minimum height=2em]
\tikzstyle{line} = [draw, -latex']
\tikzstyle{embed} = [circle, fill=blue!20, text centered, minimum height=2em]

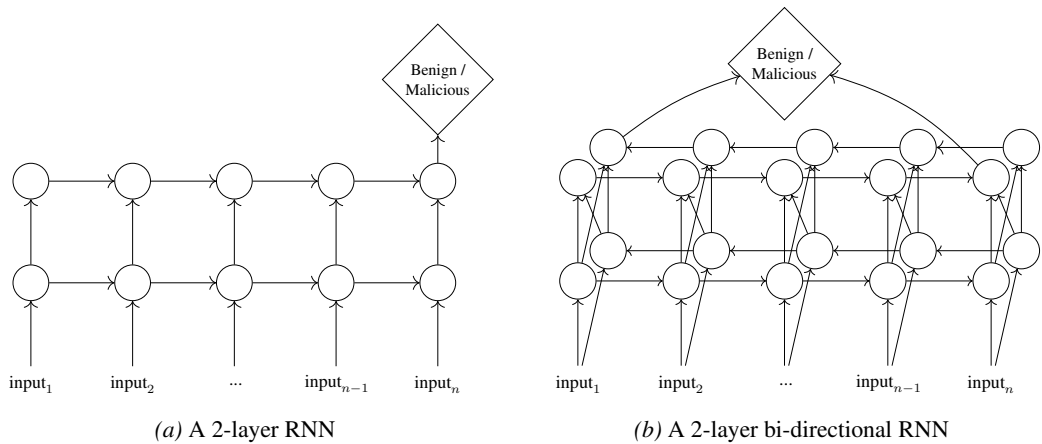
\begin{figure}[ht]
    \centering
    \begin{subfigure}[b]{0.48\textwidth}
	    \centering
        \resizebox{1.0\linewidth}{!}{\begin{tikzpicture}[node distance = 2cm, auto]

	\node [decision] (out) {\footnotesize Benign / Malicious};

    \node [hidden_layer, below of=out] (h25) {};
	\node [hidden_layer, left of=h25] (h24) {};
    \node [hidden_layer, left of=h24] (h23) {};
	\node [hidden_layer, left of=h23] (h22) {};
	\node [hidden_layer, left of=h22] (h21) {};

    \node [hidden_layer, below of=h25] (h15) {};
    \node [hidden_layer, below of=h24] (h14) {};
    \node [hidden_layer, below of=h23] (h13) {};
	\node [hidden_layer, below of=h22] (h12) {};
	\node [hidden_layer, below of=h21] (h11) {};

	\node [input, below of=h11] (b1) {$\text{input}_1$};
	\node [input, below of=h12] (b2) {$\text{input}_2$};
    \node [input, below of=h13] (b3) {...};
    \node [input, below of=h14] (b4) {$\text{input}_{n-1}$};
    \node [input, below of=h15] (b5) {$\text{input}_n$};
    
    \draw [-{>[scale=1.5]}] (b1) -- (h11);
    \draw [-{>[scale=1.5]}] (b2) to (h12);
    \draw [-{>[scale=1.5]}] (b3) -- (h13);
    \draw [-{>[scale=1.5]}] (b4) -- (h14);
    \draw [-{>[scale=1.5]}] (b5) -- (h15);
    
	\draw [-{>[scale=1.5]}] (h11) -- (h21);
    \draw [-{>[scale=1.5]}] (h12) -- (h22);
    \draw [-{>[scale=1.5]}] (h13) -- (h23);
    \draw [-{>[scale=1.5]}] (h14) -- (h24);
    \draw [-{>[scale=1.5]}] (h15) -- (h25);
    
    \draw [-{>[scale=1.5]}] (h11) -- (h12);
    \draw [-{>[scale=1.5]}] (h12) -- (h13);
    \draw [-{>[scale=1.5]}] (h13) -- (h14);
    \draw [-{>[scale=1.5]}] (h14) -- (h15);
    \draw [-{>[scale=1.5]}] (h21) -- (h22);
    \draw [-{>[scale=1.5]}] (h22) -- (h23);
    \draw [-{>[scale=1.5]}] (h23) -- (h24);
    \draw [-{>[scale=1.5]}] (h24) -- (h25);

    \draw [-{>[scale=1.5]}] (h25) -- (out);
    
\end{tikzpicture}}
        \caption{A 2-layer RNN}
        \label{fig:rnn}
    \end{subfigure}
    \hfill
    \begin{subfigure}[b]{0.48\textwidth}
	    \centering
        \resizebox{1.0\linewidth}{!}{\begin{tikzpicture}[node distance = 2cm, auto]

	\node [decision] (out) {\footnotesize Benign / Malicious};

    \node [hidden_layer, below=0.75cm of out] (h23) {};
	\node [hidden_layer, left of=h23] (h22) {};
	\node [hidden_layer, left of=h22] (h21) {};
	\node [hidden_layer, right of=h23] (h24) {};
    \node [hidden_layer, right of=h24] (h25) {};

    

    \node [hidden_layer, below left of=h25] (h04b) {};
    \node [hidden_layer, right of=h04b] (h05b) {};
    \node [hidden_layer, left of=h04b] (h03b) {};
	\node [hidden_layer, left of=h03b] (h02b) {};
	\node [hidden_layer, left of=h02b] (h01b) {};
    
    \node [hidden_layer, above of=h04b] (h24b) {};
    \node [hidden_layer, above of=h05b] (h25b) {};
    \node [hidden_layer, above of=h03b] (h23b) {};
	\node [hidden_layer, above of=h02b] (h22b) {};
	\node [hidden_layer, above of=h01b] (h21b) {};
    
    \node [hidden_layer, below of=h25] (h05) {};
    \node [hidden_layer, below of=h24] (h04) {};
    \node [hidden_layer, below of=h23] (h03) {};
	\node [hidden_layer, below of=h22] (h02) {};
	\node [hidden_layer, below of=h21] (h01) {};

    \node [input, below of=h05] (e5) {$\text{input}_n$};
    \node [input, below of=h04] (e4) {$\text{input}_{n-1}$};
    \node [input, below of=h03] (e3) {...};
	\node [input, below of=h02] (e2) {$\text{input}_2$};
	\node [input, below of=h01] (e1) {$\text{input}_1$};

    \draw [-{>[scale=1.2]}] (e1) -- (h01);
    \draw [-{>[scale=1.2]}] (e2) -- (h02);
    \draw [-{>[scale=1.2]}] (e3) -- (h03);
    \draw [-{>[scale=1.2]}] (e4) -- (h04);
    \draw [-{>[scale=1.2]}] (e5) -- (h05);
    \draw [-{>[scale=1.2]}] (e1) -- (h01b);
    \draw [-{>[scale=1.2]}] (e2) -- (h02b);
    \draw [-{>[scale=1.2]}] (e3) -- (h03b);
    \draw [-{>[scale=1.2]}] (e4) -- (h04b);
    \draw [-{>[scale=1.2]}] (e5) -- (h05b);
    
    \draw [-{>[scale=1.1]}] (h01) -- (h21);
    \draw [-{>[scale=1.1]}] (h02) -- (h22);
    \draw [-{>[scale=1.1]}] (h03) -- (h23);
    \draw [-{>[scale=1.1]}] (h04) -- (h24);
    \draw [-{>[scale=1.1]}] (h05) -- (h25);
    \draw [-{>[scale=1.1]}] (h01b) -- (h21b);
    \draw [-{>[scale=1.1]}] (h02b) -- (h22b);
    \draw [-{>[scale=1.1]}] (h03b) -- (h23b);
    \draw [-{>[scale=1.1]}] (h04b) -- (h24b);
    \draw [-{>[scale=1.1]}] (h05b) -- (h25b);
    \draw [-{>[scale=1.1]}] (h01) -- (h21b);
    \draw [-{>[scale=1.1]}] (h02) -- (h22b);
    \draw [-{>[scale=1.1]}] (h03) -- (h23b);
    \draw [-{>[scale=1.1]}] (h04) -- (h24b);
    \draw [-{>[scale=1.1]}] (h05) -- (h25b);
    \draw [-{>[scale=1.1]}] (h01b) -- (h21);
    \draw [-{>[scale=1.1]}] (h02b) -- (h22);
    \draw [-{>[scale=1.1]}] (h03b) -- (h23);
    \draw [-{>[scale=1.1]}] (h04b) -- (h24);
    \draw [-{>[scale=1.1]}] (h05b) -- (h25);

    \draw [-{>[scale=1.1]}] (h01) -- (h02);
    \draw [-{>[scale=1.1]}] (h02) -- (h03);
    \draw [-{>[scale=1.1]}] (h03) -- (h04);
    \draw [-{>[scale=1.1]}] (h04) -- (h05);
    \draw [-{>[scale=1.1]}] (h21) -- (h22);
    \draw [-{>[scale=1.1]}] (h22) -- (h23);
    \draw [-{>[scale=1.1]}] (h23) -- (h24);
    \draw [-{>[scale=1.1]}] (h24) -- (h25);
    \draw [-{>[scale=1.1]}] (h02b) -- (h01b);
    \draw [-{>[scale=1.1]}] (h03b) -- (h02b);
    \draw [-{>[scale=1.1]}] (h04b) -- (h03b);
    \draw [-{>[scale=1.1]}] (h05b) -- (h04b);
    \draw [-{>[scale=1.1]}] (h22b) -- (h21b);
    \draw [-{>[scale=1.1]}] (h23b) -- (h22b);
    \draw [-{>[scale=1.1]}] (h24b) -- (h23b);
    \draw [-{>[scale=1.5]}] (h25b) -- (h24b);

    
    \draw [-{>[scale=1.5]}] (h25) to [bend right=15] (out);
    \draw [-{>[scale=1.5]}] (h21b) to [bend left=10] (out);

\end{tikzpicture}}
        \caption{A 2-layer bi-directional RNN}
        \label{fig:rnn_bidir}
    \end{subfigure}
    \caption{Two simpler RNN architectures for classifying a sequence. In each case the rows of the diagram represent the same neuron being applied on inputs at different time steps. The input includes the bottom up input (previous layer) as well as the previous time step (which is implicitly zero when not present).}
    \label{fig:rnn_examples}
\end{figure}

In the Deep Learning literature the Long Short Term Memory (LSTM) unit \cite{Hochreiter1997} has also helped to overcome a number of difficulties in training the recurrent connections themselves, especially in combination with recent advances in gradient-based training and weight initialization. Training works by extending back-propagation "through time" \cite{Werbos1988}, which amounts to unrolling the sequence of input and output transitions over the course of the sequence (i.e., weight sharing across time). This produces a directed acyclic graph, on which normal back-propagation can be applied. Two examples of this are given in \autoref{fig:rnn_examples}, where the neurons in each column all share the same weights. Back-propagation can then be done normally on this unrolled architecture, and the shared weights will be updated based on the average gradient for each time the shared weight was used. This also means that any of the architecture types discussed in \autoref{sec:neural_nets} can be used with RNNs, either before, after, or in-between recurrent stages, and trained jointly with the rest of the network. 

\textcite{Kolosnjaji2016} have exploited the flexibility of neural networks to combine LSTMs with CNNs for malware classification based on API call sequences. The architecture combination allows the CNNs to learn to recognize small groupings of co-occurring API calls, and the LSTM portion allows the information from multiple co-occurrences to be captured through the whole call trace to inform a decision, and were found to out-perform HMMs by 14 to 42 percentage points. Similar work was done by \textcite{Tobiyama2016}. 

Using just LSTMs, \textcite{raff2017peheader} instead looked at byte sequences and were able to show that LSTMs could learn to process the bytes of the PE header sequentially to arrive at accurate classifications. They also used an \textit{attention mechanism} to show that the LSTM learned to find the same features as a domain knowledge approach learned to use when the features were manually extracted. The purpose of an attention mechanism is to mimic the human ability to focus on only what is important, and ignore or down-weight extraneous information. This has become a tool often used in Machine Translation, and offers a more interpretable component to a neural network. 

CNNs to process the raw bytes of a file where first introduced by \textcite{MalConv}, which treated the raw bytes of a executable file as a 2 million byte long sequence. Their work found that many of the best practices for building neural networks for image, signal, and natural language processing did not carry over to learning from raw bytes. Notably this required using a shallower and wider network (rather than deep and narrow), and the abandonment of layer normalization techniques like Batch-Norm\cite{Ioffe2015}. \textcite{Krcal2018} expanded this work on their own corpus, but also compared with analyst-derived features used by Avast for their AV product. In doing so they found the approach was competitive with their classical AV, but that combining the features learned by the CNN with those their analysts developed had the best accuracy. This indicates that the CNN approach is learning to detect features that were overlooked by the expert analysts. Otherwise the features would be redundant, and their combination should have no impact. 

The ability of CNNs and RNNs to handle variable length sequences makes them an attractive tool for malware classification, however they have yet to receive significant application to that task. Part of this is likely due to the added computational burden they bring. Regular neural network algorithms often require the use of GPUs to train for days at a time. RNNs exacerbate this situation with more computations and a reduction in the parallelism of the problem, making it challenging to scale out training across multiple nodes. It is also difficult to train RNNs for sequences longer than a few thousand time steps, which is easily exceeded by call traces, byte sequences, and entropy measurements. It is likely that intelligent combinations of RNNs with other architectures or advancements in training efficiency will be needed before wide use is seen.

\subsection{Haar Wavelet Transform}

For numeric sequences, such as the windowed entropy of a file over time, wavelets provide a versatile tool for tackling the problem in a number of ways. Wavelets have long been used to perform machine learning over sequences, and are particularly popular for signal processing tasks \cite{Chaovalit:2011:DWT:1883612.1883613}.  At a high level, wavelets simply aim to represent an underlying signal $f(t)$ sampled at $N_f$ points using a combination of waves added together. A wave is just a function over time that starts and ends at zero, and goes above and below zero in the middle. The seminal Fast Fourier Transform is one type of continuous wavelet that represents a function as a combination of sine and cosine waves. 

For malware classification, the Haar wavelet is becoming increasingly popular and has been used in a number of different ways \cite{DBLP:conf/flairs/WojnowiczCW16,Han2015,Baysa2013,Sorokin2011}. 
Of all possible wavelets, the Haar wavlet is discrete and the simplest possible wavelet. The \textit{Haar wavelet} $\psi^{\text{}}(t)$ is defined in \eqref{eq:haar_mother}, and has non-zero vales in the range $[0, 1)$. It is positive in the first half of the range, and then negative in the second. 

\begin{equation}\label{eq:haar_mother}
\psi^{\text{}}(t) = \begin{cases}
  1  & \text{if } t \in [0, \tfrac{1}{2}) \\
  -1 & \text{if } t \in [\tfrac{1}{2}, 1) \\
  0  & \text{otherwise}
\end{cases}
\end{equation}

\begin{equation} \label{eq:haar_func}
\psi_{j,k}(t)=2^{j/2} \psi^{\text{}}(2^j t-k)
\end{equation}

Since an arbitrary function $f(x)$ cannot in general be approximated by adding combinations of \eqref{eq:haar_mother}, we need another function that moves the location of the wavelet. Thus the \textit{Haar function} is defined by \eqref{eq:haar_func}, and is used in a hierarchy of levels to approximate functions. Both $j$ and $k$ are integers, where $j$ is the level of the hierarchy and $k \in [0, 2^j-1]$ shifts the location of the wavelet. The smallest levels of the hierarchy ($j=0$) represent course information over large portions of the sequence, and the largest values ($j \to \infty $) represent the smallest details that apply only to specific time steps $t$. 

\begin{equation} \label{eq:haar_approx}
f(t) = c_0 + \sum_{j=0}^{\log_2 N_f} \sum_{k=0}^{2^j-1} c_{j, k} \cdot \psi_{j, k}(t)
\end{equation}

We can now represent any function $f(t)$  using \eqref{eq:haar_approx}. However, in practice this only works for sequences that are a power of 2 in length, so truncation to the smallest power of two or padding the sequence with zeros is often necessary.  Computing the Haar wavelet transform then determines the values for the coefficients $c_{j, k} \forall j, k$. The discrete Haar transform takes only $O(N_f)$ time to compute, making it a viable tool for longer sequences \cite{Walker2008}. 

A common use case of the Haar transform is to de-noise signal data, and has been used in that context for malware classification \cite{Shafiq2009}. The intuition being that the higher level signal comes from the coefficients from the earlier levels, so we can remove noise by setting $c_{j, k} = 0 \forall j \geq z$, for some threshold $z$. Once set to zero, we can then re-construct the original signal which should now have less noise present. While the de-noising application is fairly straightforward and is a standard use case for wavelets, once we have this representation there exist a number of possibilities for comparing sequences. 

\subsubsection{Haar Distance} \textcite{Struzik:1999:HWT:645803.669368} propose comparing time series by defining a distance over the coefficients $c_{j,k}$. First they define a correlation between two Haar transforms as given in \eqref{eq:haar_correlation}, where the Kronecker delta function $\delta(j_a, k_a, j_b, k_b)$ returns 1 iff $j_a = j_b$ and $k_a = k_b$, otherwise it returns zero. Using this a normalized distance between two sequences is defined by \eqref{eq:haar_distance}. While this would allow for a nearest neighbor classifier to be used, this specific approach has not yet been used for malware classification. 

\begin{equation}\label{eq:haar_correlation}
C(f_a, f_b) = \sum_{j_a=0}^{\log_2 N_{f_a}} \sum_{k_a = 0}^{2^{j_a}-1} \sum_{j_b=0}^{\log_2 N_{f_b}} \sum_{k_b = 0}^{2^{j_b}-1} c_{j_a, k_a} c_{j_b, k_b} \delta(j_a, k_a, j_b, k_b)
\end{equation}

\begin{equation}\label{eq:haar_distance}
D_{\text{Haar}}(f_a, f_b) = - \log  \left| \frac{C(f_a, f_b)}{\sqrt{C(f_a, f_a) \cdot C(f_b, f_b)}} \right| 
\end{equation}

\subsubsection{Haar Energy Features} Another approach is to extract a fixed-length representation from the Haar transform, and use those features as inputs into a classifier \cite{Pati1993}. \textcite{DBLP:conf/flairs/WojnowiczCW16} used this approach on entropy sequences, and found them to be moderately effective independently, but not sufficient for a complete system. For a sequence of length $N_f$, we can compute $\log_2 N_f$ energy levels according to equation \eqref{eq:haar_energy_j}. In the case of \textcite{DBLP:conf/flairs/WojnowiczCW16}, they decided to use all possible levels and handled different length sequences by building a new classifier for every different power of two. 

\begin{equation} \label{eq:haar_energy_j}
\text{Energy}_j  = \sum_{k=1}^{2^{j-1}} (c_{j, k})^2
\end{equation}

\subsection{Dynamic Programming} \label{sec:dynamic_programing} 
A method of comparing numeric sequences that has become increasingly common in the literature is to perform distance computations using dynamic programming methods such as the Levenshtein distance \cite{levenshtein1966binary,Bellman1954}. While the use of so-called edit-distances could be applied directly to raw binaries, a run-time complexity of $O(N^2)$, where $N$ is the length of a binary or entropy sequence, is prohibitive. The Haar transform has been used to make such approaches practical. The Haar transform is used to discretize the numeric entropy sequence into bins of varying size (resulting in a shorter sequence), which are then used in the dynamic programming solution \cite{Han2015,Baysa2013,Sorokin2011,Shanmugam:2013:SSD:2509225.2509226}. While similar at a high level, these approaches all have a number of differences and details that do not lend themselves to a compact summary. 

Instead we summarize a similar dynamic programming distance for sequences called \textit{Dynamic Time Warping} (DTW) \cite{Berndt:1994:UDT:3000850.3000887}. DTW works directly upon numeric time series, and does not need the discretization provided by the Haar Transform (though the wavelets could be used to pre-process the signal before computing DTW).  This method has received minimal investigation for malware classification \cite{Naval:2014:EWB:2659651.2659737}, but has been used for related tasks such as estimating the prevalence or malware infection \cite{Kang:2016:EMD:2835776.2835834} and detecting botnets \cite{Thakur2012}. Its description is also representative of a simpler common ground between the prior work in this area \cite{Han2015,Baysa2013,Sorokin2011,Shanmugam:2013:SSD:2509225.2509226}. 

Given two sequences $f_a(t)$ and $f_b(t)$ of lengths $N_{f_a}$ and $N_{f_b}$ respectively, DTW attempts to find a continuous path of point-wise distance computations that minimizes the total distance. Doing so requires finding a sequence of dilations and contractions of one sequence to fit the other, such that it maximizes the similarity between their shapes. In other words, DTW attempts to measure the distance between two sequences based on overall shape, and ignoring any local contractions or elongations of a sequence. To do so, we define the distance using the recursive  equation \eqref{eq:dtw_distance}, where $\text{DTW}(f_a, f_b) = \text{DTW}(N_{f_a}, N_{f_b})$. This equation can be solved using dynamic programming, and is similar to the Levenshtein style distances that have been more prevalent in the malware classification literature.  

\begin{equation}\label{eq:dtw_distance}
\text{DTW}(i, j) = \left( f_a(i)-f_b(j) \right)^2 +
\begin{cases}
0 & \text{if } i = j = 1 \\
\min \begin{cases} \text{DTW}(i-1, j) \\ \text{DTW}(i-1, j-1) \\ \text{DTW}(i, j-1) \end{cases} & \text{otherwise}
\end{cases}
\end{equation}

Like many other dynamic programming methods, DTW takes $O(N_{f_a} N_{f_b})$ time to compute and has the disadvantage that it is not a true distance metric, as it does not obey the triangle inequality. However, the popularity of DTW in other domains warrants its consideration for malware classification, especially as an existing body of literature addresses many of the known problems and challenges. In particular there are multiple methods of speeding up the DTW computation and retrieval, including the construction of indexes \cite{Keogh2005}, a fast $O(\max(N_{f_a}, N_{f_b}))$ approximation \cite{Salvador:2007:TAD:1367985.1367993}, and a definition for a DTW centroid \cite{Petitjean2014}. There even exist methods of learning constraint costs to modify the DTW, which can improve the accuracy of the constrained DTW \cite{Ratanamahatana2004}. This could simplify the manually constructed costs and constraints used in existing malware research \cite{Han2015,Baysa2013,Sorokin2011}. 

\section{Graph Algorithms} \label{sec:graphs}

While representing information as a sequence reduces the gap between abstraction and the true nature of the data, this is still a simplification in many instances. For example, while a binary may be one long sequence of bytes, the order in which the bytes are evaluated and accessed may be non-linear. A yet richer representation is as a graph $G$ of vertices $V$ and edges $E$. Graph analysis and algorithms have been widely studied and used in many domains, and malware classification is no exception. Such techniques have been used most commonly for representing assembly\cite{Alam2015212,Anderson2011,Hashemi2016} and API and system calls \cite{Elhadi2014} collected from dynamic analysis. While these two feature types have already seen use as sequences and with classical machine learning, graphs have also been used for features not well represented in other forms, like mapping the relationship between a malware \textit{per se} and the files it creates or accesses \cite{export:193769}. 

Similar to \autoref{sec:ml_seq}, we will review the common graph-based approaches that have been used for malware classification. While the appropriateness of a graph representation has been widely recognized in prior work, little has been done to fully exploit such representations. Most of the prior work in graphs for malware classification use either graph matching, or attempt to construct more meaningful feature vector representations that will allow us to use the wide breadth of machine learning methods in \autoref{sec:ml_methods}. While there exists a rich diversity in what may be done with graphs, a full survey of graph methods is beyond the scope of this work. Instead we will review the two high-level approaches that have been used in the malware classification literature. 

\subsection{Graphs to Feature Vectors}

Many of the techniques for directly comparing graphs are computationally demanding, often requiring approximations or simplifications to be made. For this reason it is often desirable to convert a graph representation to a fixed length feature vector. Once converted, the classical machine learning methods discussed in \autoref{sec:ml_methods} can be used.  The simplest approach is to manually craft features about nodes in the graph, and then use those for classification (e.g., as used by \textcite{Kwon:2015:DEI:2810103.2813724} ). 

\subsubsection{Graph Features and Descriptors}

There exist naive ways to create a feature vector from a graph, 
such as flattening the $n$ by $n$ adjacency matrix into a vector of length $n^2$ \cite{Eskandari2012}. But such approaches are often impractical, causing excessively high dimensional spaces (making learning more challenging) and relying on extremely sparse graphs. Another simple approach to creating feature vectors from graphs is to derive informative statistics about the graph itself. Such features could include the number of vertices or degrees with a given label, the average number of in/outbound edges for a vertex, and various other statistics. This approach was used by \textcite{Jang:2014:MMC:2567948.2579364} over the system call graph of a binary. This approach has the benefit of being quick to implement and apply, but does not take full advantage of the rich information that can be stored within a graph. It also requires the developer to have some notion about which statistics will be informative to their problem. 

These kinds of approaches are often not sufficient in practice, but are easier to apply and generally not compute intensive. Outside of malware classification, machine learning combined with these simple feature types can be used to accelerate other approaches as a kind of low-pass filter, pre-classifying items as likely or unlikely to be related \cite{Lazarescu:2000:GMF:645889.673411}. This filtering can be used to reduce the compute time of the other approaches we will discuss in this section, which are often more compute intensive. 

\subsubsection{Graph Embedding}

A less labor-intensive means of converting a graph into a feature vector is to perform some type of embedding, where the graph is converted to a fixed-length feature vector through some decomposition \cite{Luo2003}. This is often done by taking the adjacency matrix representation of a graph, and performing an eigenvalue decomposition (or singular value decomposition (SVD)), and using the eigenvalues
as the feature vectors. This captures information about the graph as a whole, and does not require manually specifying items of interest. 

These embeddings can be used directly for machine learning, or can also be used as a first step toward other methods discussed in this section, such as graph matching in \autoref{sec:graph_dist_match} \cite{Luo2001}. \textcite{Hashemi2016} used this strategy with assembly code, representing an edge in the graph every time one instruction followed another and weighting the edges by the number of times the instruction pairing occurred. \textcite{Anderson2011} has also used this strategy for assembly sequences in combination with other feature transforms of a graph. A unique approach by \textcite{Slawinski2019} defined graphs over the extract abstract syntax trees of functions, and used a PageRank based embedding to better capture information from statically extracted features. Using just the static features in a fixed length feature vector they obtained 89.19\% accuracy, which increased to 98.28\% when leveraging the graph weighted embedding~\cite{Slawinski2019}.

\subsection{Graph Matching} \label{sec:graph_dist_match}

The most prevalent method in use for malware classification is the concept of \textit{graph matching}. Given two graphs $G_1$ and $G_2$, in the case of this work, the goal is to derive some distance function that will give us a measure of how close $G_2$ is to $G_1$. This can be done with all types of graphs, and the methods used may change with the graph size and type. 

The general strategy to create this distance measure is to determine the amount of correspondence between the graphs, often by determining how to transform one into the other. There are various strategies to using graph matching as well, with some works even defining their own matching algorithms \cite{HaoranGuo2010}, though more simple heuristics are more common \cite{182795}. One method is to create templates of what malware looks like, and match to that template \cite{HaoranGuo2010}. Graph matching is also popular for malware family classification, where such systems are often designed for low false positive rates \cite{Park:2011:DCM:1966913.1966986}. Graph matching can also be used for normal nearest neighbor type classification against a large database, though this is often avoided due to computational expense \cite{Hu:2009:LMI:1653662.1653736}. 

\subsubsection{Graph Edit Distance} One common approach to graph matching is to create a graph edit distance. Computing the exact graph edit distance is generally NP-complete, and so various different approximations are used in practice. Such approximations are often done via dynamic programming strategies, as discussed in \autoref{sec:dynamic_programing}. These approximations can still be too slow to use for larger corpora, which has also spurred the use of indexing structures and other optimizations to accelerate queries for similar binaries\cite{Hu:2009:LMI:1653662.1653736}.

\textcite{Park:2010:FMC:1852666.1852716} determined distances by finding the maximal common sub-graph between $G_1$ and $G_2$, and returned the edit similarity as the cardinality of the sub-graph over the maximum cardinality of $G_1$ and $G_2$. Their graph also used vertex and edge labels depending on the arguments used. Related work has used this common sub-graph approach to build systems with the goal of having minimal false positive rates \cite{Park:2011:DCM:1966913.1966986}, producing a kind of system-call "signature" for finding malware families.  

\textcite{Elhadi2014} used graph matching on labeled graphs derived from both the system call traces and the operating system resources used by each call. The edges and vertices in the graph had different labels depending on the whether the vertices came from each respective group. 

\subsection{Other Approaches}

There are many ways to perform similarity comparisons between graphs, not all of which can be described as a variant of graph matching. \textcite{Lee:2010:DMM:1774088.1774505} used a metric similar to the Jaccard distance, by having a finite space of possible vertices they were able to take the intersection of the graphs' edges over the union of the edges. Part of the motivation of this was faster compute time so that the system could be practical for real use. Their approach was also unique in that they used call traces from static code analysis, rather than dynamic execution. However, this did necessitate that their approach use unpacking software before code analysis. 

Graphs also need not be the method by which similarity is done, but can still be an integral component of the system. \textcite{Eskandari2013} used an extensive graph-based approach to process API and system calls from both static and dynamic analysis, and used the graph structure with node classification to infer what a dynamic analysis would have looked like given only the static analysis of the binary. This was to obtain the speed benefits of static analysis (at runtime), while retaining the accuracy benefits of dynamic analysis. Yet their system used a simple fixed length feature vector for the final determination of maliciousness.

\section{Evaluation and Metrics} \label{sec:evaluation}

We have now extensively considered the many predictive approaches that have been applied to malware classification. While such techniques are often the most exciting or interesting part of an approach, it is often done without fully considering how such systems should be evaluated. Despite its importance, evaluation metrics and choices are a commonly overlooked part of the machine learning process. Even more so, such choices often do not consider the evaluation of a system as a whole. Most works will simply use overall accuracy or AUC, which are described with some other less frequently used metrics in \autoref{tbl:common_eval_metrics}. 

While maximizing the performance of these metrics can be revealing and informative on its own, it is often done without necessarily justifying why these metrics are the ones to be used, or considering other explicit or implied goals. Each possible metric will be biased towards certain properties, which may impact what model appears to work best even, if it performs worse in practical use \cite{citeulike:12882259}.  Thought should be given to system  constraints, the metric of performance, and the specific use case for deployment. The latter of these concerns should in general drive the process as a whole, informing us as to which constraints we can computerize and which scoring methods most closely reflect our needs. 

Early work by \textcite{Marx2000} developed a thorough set of guidelines and processes for evaluating an AV product, from ease of user by customers to developing and weighting a testing set, and how to perform testing for different types of malware. The crux of their argument is the necessity for evaluation to reflect real world use. It was recognized then that accurately reflecting real world use of malware detectors is anything but trivial, and (we argue) has only gotten more challenging over time. 

\begin{table}[ht]
\centering
\caption{A few metrics that have been used within the malware classification literature. Most works have used only accuracy and AUC.}
\label{tbl:common_eval_metrics}
\adjustbox{max width=\linewidth}{
\begin{tabular}{>{\kern-\tabcolsep}lm{10cm}<{\kern-\tabcolsep}}
\toprule
\multicolumn{1}{c}{Metric} & \multicolumn{1}{c}{Description}                                                                                                              \\ \midrule
Accuracy                   & The number of data points classified correctly, divided by the total number of data points                                                     \\
\rowcolor{gray!20}
Balanced Accuracy          & Same as accuracy, except each class has equal weight to the final result. See \cite{Brodersen:2010:BAP:1904935.1905533}.                                                                     \\
Precision                  & The number of true positives divided by the true positives and false positives                                                               \\
\rowcolor{gray!20}
Recall                     & The number of true positives divided by the true positives and false negatives                                                               \\
AUC                        & Takes the integral of the Receiver operating characteristic curve, which is a plot of the true positive rate against the false positive rate. See \cite{Bradley1997}. \\
\rowcolor{gray!20}
F1 / F-Measure             & The harmonic mean between precision and recall.                                                                                                 \\ \bottomrule
\end{tabular}
}
\end{table}

In this section we will review a number of different scenarios and constraints that may impact how we choose to design and evaluate a malware classification system. Intrinsically related to this is the data quality issues we discussed in \autoref{sec:data}. Having high quality data not only means better models but more accurate estimation of model performance. While we have already reviewed why this is not a given, for brevity we will generally assume that the data issue has been resolved in this section. 

\subsection{Malware Detection} \label{sec:mal_detection}

The most obvious goal of malware detection is to act as a replacement for anti-virus products. In this case accuracy and AUC are both acceptable target metrics that are widely used in machine learning. This does not necessarily make them the best metrics when we consider the user of such a system. Numerous works have emphasized the importance of having low false-positives in this scenario 
\cite{Ferrand2016,Masud2008,Alazab:2011:ZMD:2483628.2483648,Zhou:2008:MDU:1456377.1456393}. 
This is because excessive false positives will be aggravating for the end user, who presumes their desirable goodware applications will continue to work without issue when using an anti-virus. If required to add applications to a white-list too frequently, they will get frustrated and stop using the product. This leaves them exposed to malware, and the accuracy of the system becomes irrelevant. 
While no consensus has been reached as to an acceptable false positive rate, most work that emphasizes the issue achieve rates between 1\% and 0.1\% \cite{Yan:2015:SYE:2714576.2714578,Shafiq2009}. Existing anti-virus products have been tested with fewer than 20 false positives out of over 1 million test files \cite{avtest_product_eval_2016}, though the exact details of the test set are not public. 

Another use case for malware detection is to rank a backlog of binaries for analysis or investigation. This scenario can occur for malware analysts or incident response personal, when a large number of items need to be considered and it is important to find the malware as soon as possible. The AUC metric is in fact well designed for this use case, as it can be interpreted as measuring the quality of a ranking. In this way the metric accurately aligns to the goal, in that we want to make sure analysts spend their time looking primarily at malware first. If left unordered, the chance of dealing with a benign file before all malware is dealt with increases, which takes up valuable time but does not result in any new insights. 

\subsection{Malware Family Classification} \label{sec:mal_family_classification}

When performing malware family classification, the normal procedure is to collect some number of malware families $C$, and divide the dataset into a training and testing set (or use cross validation) to evaluate the accuracy of determining the malware family for new binaries. Because this situation is not commonly needed for desktop deployment, the same run-time constraints are not often emphasized for this task. This evaluation is somewhat flawed with respect to real-life use case scenarios, as inevitably new malware families will be found that are not in the existing set. We argue that there should to be a pseudo $C+1$'th family for "not in existing families" when evaluating malware family classification. The binaries that will belong to this $C+1$'th class should also come from multiple malware families, and not be present in the training data. This will better evaluate the quality of a system with respect to situations that will be encountered during real-world usage, but does increase the design burden by requiring both a classification and anomaly detection ability. 

The importance of correctly marking a file as "not previously seen" is exemplified in forensic use cases, where a set of known malware is compared against potentially terabytes of data\cite{Roussev2010,Kornblum2006,Roussev2012}. Similarity preserving hash functions, which have low false positive (and recall) rates, are often used in this scenario. 

If we accept the need for a "not previously seen" category, it is also worth considering if benign files should be included in the evaluation. Normally malware family classification is done under the assumption that all files \textit{are} malicious. In practical use, errors will occur --- and so it is likely a benign file will occasionally be given. It seems reasonable that the "best" option (given that the benign file was already mislabeled as malicious) is to mark the benign file as "not previously seen". This is an issue we have not yet seen discussed or included in evaluations.

We also note that while a rich diversity of metrics have been used with binary classification problems, such as 
accuracy, AUC, $\text{F}_1$, Matthews correlation coefficient, etc., most works on malware family classification simply use accuracy. This is not necessarily a good choice, as the distribution of malware families is not uniform.  While balanced accuracy is one alternative, it can also pose a problem with very rare malware families. These families will be harder to classify yet have a more significant impact on the balanced accuracy. There is also a question as to how many malware families should be used when evaluating. Others have also proposed that malware should be instead grouped by behavior or effects, given the non-standardized and inconsistent nature of malware family labeling \cite{Bailey2007}.

\subsection{Training and Inference Constraints}

The most common type of constraints that are considered are run time requirements. In particular many works have been concerned with real-time inference \cite{large-scale-malware-classification-using-random-projections-and-neural-networks,Alam2015212,Khan2010}, which usually implies a strict or moderate limit on memory use as well. This scenario makes perfect sense from a deployment perspective, for systems that would act in the stead of anti-virus products or network appliances that would inspect live traffic. If such systems were to impose too significant a latency on application start up or network connections, people would stop using the product due to the loss in productivity and aggravation. If this real-time requirement is violated, accuracy becomes irrelevant because the solution is not used. This situation would also discourage us from considering the full breadth of dynamic features, as certain combinations may prove too compute intensive to be practical. While many works report sub-second classification time per datum, there appears to be no current consensus on where the threshold for "real-time" is for malware classification, and most works simply emphasize that their solutions execute quickly. 

Another consideration that is less frequently addressed is training time and scalability, particularly as corporate malware collections are on the order of hundreds of millions, if not billions, of samples \cite{tagkey2014iv}. In particular it would be ideal for a malware training system to require only one pass over the training data, as data access often becomes the bottleneck at larger scales \cite{Wojnowicz2016}. 
Not only does reducing the training time save money (as less resources are needed), but it also allows for tackling the problem of concept-drift through change-detection \cite{Gama2004,Baena-Garcia2006,Bo-HengChen2014}. This is a common method of dealing with the difficulties of adapting a model to a changing concept. Instead one uses a change detection algorithm to determine when the concept has drifted significantly enough that accuracy may begin to decrease. At that time one then simply re-trains the model on the newest data, and switches all old models to the most recent and up-to-date version. Though an important consideration is that older data may still be useful to train on, making it necessary to balance between new data and older (yet still informative and representative) data. We refer the reader to \cite{Gama2014} for a survey of many approaches to change detection. In practice the change detection may not be needed if one instead trains at a fixed interval, say every few months. We are not aware of any work that quantifies this problem on a large dataset over many time frames.  

\subsection{Specialized Needs}

In a conversation about the constraints and metrics that must be met for a malware classification system, it is also important to discuss scenarios with specialized needs. These may be uncommon deployments or scenarios where a system designed for the majority of potential users does not satisfy important requirements. By the nature of such a scenario, we cannot enumerate all possible specialized needs. Instead we present an example scenario that has had some investigation, and how that may impact the measures of success. 

A particular variant of the malware detection problem is to detect specific types of malware, such as ones that actively try to avoid detection\cite{Stolfo2007,184519}. This can be important for researchers and analysts who wish to track more sophisticated threats, or desire to study a particular type of malware. If such binaries are going to be manually analyzed afterwards, we want to make sure that selected binaries are worth an analyst's time, and so a high precision is required from the system. It would also be appropriate to evaluate the precision at multiple thresholds, reflecting the potential capacity to analyze such binaries on the part of a team that is abnormally busy, under normal workload, or blessed with excess availability. Another way this may be desirable is based on the type of infrastructure that needs protection. If a company's services were deployed in a cloud environment, malware that brings down a single VM may not be a significant issue, as one can easily provision and replace the infected VM. However, malware that exfiltrates data on the VM to a third party may cause the loss of proprietary or confidential information, and thus be a heightened concern. In this case we want a malware evaluation model adjusted to the type of exploits which can cause the most damage to a particular entity and environment. 

\subsection{Evaluating Over Time}

It is also important to evaluate a system based on the longevity of the model's utility. That is to say, we may desire a model that is more robust to concept drift, perhaps at some cost in terms of another metric. This would be important for any system that in some way has limited Internet connectivity, making it expensive (or impossible) to update the model over time. This does not necessarily prevent malware from trying to infect the device when online, or from someone with physical access attempting to install malware on the device. In this case the model needs to be robust to malware over longer periods of time to thwart such efforts until an update of the model can be completed. In our view, \textcite{Saxe2015a} is one of the more important attempts at performing such an evaluation. They used the compile date provided in the PE header to split the dataset into before and after July 31st, 2014. Under this scenario their malware detection rate dropped from 95.2\% to 67.7\% for a fixed false positive rate of 0.1\%. 

This dramatic drop shows the importance of considering time in evaluations, which is not a common practice. One issue is that the compile date in the PE header is easily modified, and so malware authors can alter the value seen. File first-seen date may be a usable alternative source for this information, but is necessarily less precise. Performing evaluations split in time like this also requires significant data from a wide breadth of time. This exacerbates the need for good benign data mentioned in \autoref{sec:data}. Not addressed in \textcite{Saxe2015a}, but also worth considering, is evaluating the performance on the test set as a function of time --- which would allow one to more precisely characterize the longevity of generalizing information. 

The EMBER dataset follows this evaluation protocol, with date-first-seen information provided by an external source rather than the time stamp of the file \cite{Anderson2018}. This avoids the problems caused if the malware lies about its creation date, but has less precision. Valuable information could be obtained by doing a comparative study between these different types of date information, seeing how well they correlate, and how the choice impacts results. 

There are still many important questions related to a malware model's performance over time that have not been answered. Most notably, for how long is a model effective? What are the important factors to a model's longevity (i.e., is data or the model type more or less important)? How old can training data be before it becomes uninformative or even detrimental to training? Presuming that such a threshold for training data exists, do benign and malicious data have a different  "shelf-life" in terms of utility for training?

\subsection{Evaluating Under Adversarial Attack}

Machine Learning models are generally susceptible to adversarial attacks. In this context, an adversarial attack against a machine learning model means that an abstract adversary is able to modify an input such that it has not meaningfully changed, yet a machine learning model will run an increased risk of misclassifying the modified example. We refer the reader to \textcite{Biggio2017} for a review of the history of adversarial machine learning and how it is normally performed. 

For malware classification, we have a real live adversary (the malware's author) and the use of adversarial machine learning techniques can become a new tool for the malware author to avoid detection. As such the problem of attack and defending malware classifiers against such attacks is an important area for study, and potentially included as part of the evaluation metrics \cite{Fleshman2018}. The application of these techniques to the malware space is not trivial, however. Normally an input crafted by an adversary will modify its input parameters over a continuous space of values, all of which are valid inputs. For example, in an image the pixel values will change to other adjacent pixel values. However binaries can't be altered in the same way, and changing arbitrary bytes can be destructive to the executables' function. 

\textcite{Anderson2017a} showed one of the first attack methods for arbitrary binaries against arbitrary malware detectors (including regular AV products). They did this by defining a space of possible non-destructive transformations that could be applied to the binary so that its functionality would be maintained. They then trained a reinforcement learning algorithm to learn which transforms it should apply. Importantly, by uploading a sample of files to Virus Total they found that their adversarial modifications, which have no impact on functionality and no awareness of how the AV products worked, was able to evade several of them. 

To circumvent the issue of arbitrary byte changes breaking an executable \textcite{Kolosnjaji2018} and \textcite{Kreuk2018} concurrently developed an attack that avoids this difficulty. They added an unused section to the binary, which is allowed by the PE format, and constrained all of their modifications to this lone section. This allows for inserting arbitrary byte content to trick the detector, but avoids impacting the functionality in any way. While developed as an attack against the byte-based CNNs discussed in \autoref{sec:rnn}, the approach can be leveraged against other methods as well. Recently \textcite{Fleshman2018a} proposed an approach that defeats this attack in all cases, but at a cost to accuracy. 

These recent works have operated in the threat model that the adversary (malware author) can only add features. While this threat model is reasonable, it is not perfect. This is especially true in the static analysis case, where whole file transformations like packing already exist. We expect future work to attempt new ways to add information to the static case. The dynamic case is less clear, as the malicious behavior needs to eventually run for the malware to be malicious. This is muddied further by malware detecting its environment, as discussed in \autoref{sec:dynamic_challenges}. For dynamic analysis, we expect malware authors would further invest in VM evasion techniques, and there always exists the potential for more sophisticated attempts to hide actions taken from the inspecting VM. 

While these results are all new, they represent an important area in which we must evaluate each model's accuracy both against unseen binaries, and adversarialy altered ones. This is in addition to the adversarial techniques that were developed to avoid classical AVs \cite{Poulios2015,ThomasKittel2015}. 

\section{Future Research Potential} \label{sec:future_research}

At this point we have now reviewed the current data, models, and evaluation methods used for malware classification. This includes many challenges, and some potential improvements that exist throughout the whole process. While some areas have received more and less attention than others, we note that there exist relevant techniques in machine learning that have gone almost or are completely un-applied to the problem of malware classification. We briefly review a number of these that we believe could lead to future progress. 

\subsection{Multiple Views of Malware}

Given the wide array of potential feature sources, extraction methods, representations, and models that can be used for malware classification, we could consider each of these combinations a different "view" of the malware problem. Each will be biased toward detecting different types of malware with different representations. This can be a beneficial scenario for ensembling, where multiple models are combined to form a decision. It is commonly found that ensembles work best when the members of the ensemble make uncorrelated errors, which is more likely when each model has a different "view" of what malware looks like. Despite this seemingly good fit, we have found little work in using ensemble methods to combine multiple distinct feature types. \textcite{Singh2015} used a SVM to combine the predictions of three models and found it to perform considerably better (though they did not connect that this is a form of Stacking ensembles \cite{Wolpert1992} ). However, in their experiments all three models used assembly instructions as the source feature --- reducing the potential diversity of the model. \textcite{Masud2008} used three different feature types (assembly, byte n-grams, and PE header imports), but instead combined all feature types into one large feature vector. \textcite{Eskandari2013} did some work providing a different take on looking at a binary in multiple different ways. They used API call traces features collected by both static and dynamic analysis, where both were used at training and only static analysis was used during testing. They modeled what decisions might have been made in the static version from the dynamic, so that at testing time the faster static-only analysis could be used. 

There may be many other ways to exploit the rich diversity in feature and models types that can be applied to malware classification, instead of focusing on finding the single method that performs best. For example, it is possible that different model and feature types work best for differing types of binaries. A Mixture of Experts \cite{Jacobs1991} ensemble could be built that learns to recognize which models will work best on a given binary, and having that limited subset vote on the decision. We believe this is a mostly unexplored avenue for future improvements in malware classification. 

\subsubsection{Static and Dynamic Staged Views}

While we have discussed some work that uses features collected from both static and dynamic analysis \cite{Islam2013,Damodaran2015}, we have not seen any work that considers models that are first used in a static pipelines, followed by a dynamic one if the static feature model was unsuccessful. This intentionally keeps the feature spaces disjoint, and creates a cascade reminiscent to those used in early face detection work \cite{Viola2001}. 

This approach recognizes that dynamic analysis is considerably more expensive than static analysis, both in terms of time to decision and computational resources. It is unrealistic to expect to run everything through dynamic analysis, and so a tiered  approach should be used instead. This appears to be an accepted and common practice amongst AV vendors\footnote{For example, see \url{https://cloudblogs.microsoft.com/microsoftsecure/2017/12/11/detonating-a-bad-rabbit-windows-defender-antivirus-and-layered-machine-learning-defenses/}}, but we are not aware of any published research recognizing the need for such staged approaches. 

\subsection{Dealing with Labeled Data Expense}

Given the time intensive and data gathering issues discussed in \autoref{sec:data}, further research is warranted in how to perform malware classification with minimal labeled data. In this vein, it is surprising that no work has yet been done to apply \textit{semi-supervised} learning to malware classification. Semi-Supervised learning involves building a classifier using both labeled and unlabeled data \cite{Zhu2008}. Semi-Supervised learning also provides a training time workaround for when only a few anti-virus products mark a binary as malicious, casting doubt as to its true labeling. A semi-supervised solution would be to use all the data for which we are sure of the label (no anti-virus fires, or almost all fire) as labeled training data, and the ambiguous cases as unlabeled training data. In this way we avoid poisoning the learning process with bad labels, but do not throw away potentially valuable information about the decision surface. The best type of semi-supervised algorithm to use in this domain is not obvious, and many algorithms make differing assumptions about the nature that unlabeled data and the impact it should have on the decision surface. 

If we are to use a small amount of labeled data, it is also important that we label the most informative data possible. \textit{Active Learning} is a technique that can help with this. Active Learning starts with an initial dataset, and a collection of unlabeled data. The model then can select data points for which it would like to request the label. While this framework is often used to help derive algorithms that learn more efficiently and quickly, it is also directly applicable to deciding which data points we should get labels for. The potential impact of having an active learning system was shown by \textcite{Miller:2016:RIP:2976956.2976966}, where a simulation of human labelers found that their system's detection rate could be improved from 72\% to 89\%, while maintaining a 0.5\% false positive rate. However, their approach to active labeling was heuristic, and did not leverage the full potential literature of active labeling methods. There has been little other research in this area for malware classification \cite{Moskovitch2009}, and so many questions remain. Are the features best for prediction also the best for active learning? What kinds of active learning algorithms work best for this domain? How are these methods impacted by the concept-drift of binaries over time? All of these questions, as far as the authors are aware, have yet to be explored. 

\subsection{Learning with Class Imbalance} \label{sec:class_imbalance}

Most machine learning algorithms presume equally balanced classes \cite{He2009} at training time, and thus also at testing time. Learning from imbalanced classes can naturally cause the algorithm to favor the more numerous class, but also be detrimental in failing to learn how to properly separate the classes. Class imbalance is a common problem in the malware domain \cite{Patri2017,Cross2011,Li2017,Yan:2013:EDF:2530084.2530088,Moskovitch2009a,Yan:2015:SYE:2714576.2714578}, which makes it especially important to consider the evaluation metric used for both malware detection and family classification (as mentioned in \autoref{sec:mal_detection} and \autoref{sec:mal_family_classification}). 

One way to tackle such imbalance problems is to over-sample the infrequent classes or under-sample the more populous ones. These approaches are common, but can be out-performed by a more intelligent over-sampling or under-sampling process\cite{Laurikkala:2001:IID:648155.757340}. Indeed there is an existing literature for such methods focusing on both approaches\cite{JMLR:v18:16-365}, which have seen almost no application to the malware problem. Exploring their applicability to this domain and how such methods may be adapted for the difficulties of sequence and graph based features is, as far as we are aware, an open problem area. 

Oversampling the minority class is an intuitively desirable approach, as it allows us to use the larger amount of of majority class data --- and thus more data to train on overall. However, naive oversampling can lead to overfitting \cite{PratiIICAI09,Kubat1997}. One technique to do this more intelligently is to interpolate new datums from the training distribution, and a popular algorithm SMOTE takes this approach and has many variants as well \cite{Nguyen:2011:BOI:1972030.1972031,Han:2005:BNO:2141202.2141297,Chawla2002}. This also assumes a natural fixed length feature vector, and that interpolated instances are intrinsically meaningful. This may not be the case for all malware features, and may not be readily possible for sequence or graph based approaches to malware classification. 

Under-sampling is often done intrinsically for malware detection, where the ratio of malicious to benign samples available is large. While less intuitively desirable than oversampling, smarter approaches for this exist as well. One approach is to train multiple models, with different subsets of the majority class used in each model, allowing for the creation of an ensemble\cite{Liu2009}. Other techniques also attempt to more intelligently select the sub-samples\cite{Kubat1997}. 

We are aware of relatively little work that explores the problems of class imbalance in this space. \textcite{Moskovitch2009a} investigated training set proportion's impact on differing test-set proportions, but worked under the assumption that malware will make up 10\% of seen files in a network stream, and in later work simply set the training ratio equal to this ratio \cite{Moskovitch2009}.
\textcite{raff_shwel} looked at developing an algorithm specific oversampling technique, evaluating on an assumption of equal class ratio. 

It is generally thought that malware will make the minority of samples in deployment,  which is supported by a large study on over 100 million machines which found that the number of benign singletons outnumbers malicious ones at a ratio of 80:1 \cite{Li2017}. However, they also found that this ratio changed over time. We think further study is warranted to determine how applicable this rule of thumb is. Do all institutions share the same ratio of benign to malicious, or would certain institutions or individual users who are targets for malware authors have lower ratios? Similarly, do different types of equipment (e.g., desktop computer, router, or mobile device) see differing ratios of malware? How do these rates change with geographical area? Lastly, we suspect there are a number of niche professions with special needs that will see differing distributions. For example, cyber crime investigators inspecting a recovered hard drive may expect to see a much higher ratio of malware given their targeted goals.  

\subsection{The Good, the Bad, and the Annoying}

For malware detection, we also note the the distinction of malicious versus benign may be overly strong. Some anti-virus products refer to what is known as "potentially unwanted programs"\footnote{As an example, the free anti-virus ClamAV has signatures for such software \url{https://www.clamav.net/documents/potentially-unwanted-applications-pua}} (PUP). This third class sits in a gray area between benign and malicious. While these PUP binaries do not intentionally do any harm, they may be annoying or aggravating to the user. They also can present a labeling challenge, where it is not clear on which side of the line between benign vs malicious they should fall. While one could  treat this as a three class problem, it would be more accurate to apply ordinal regression to alleviate the issue. Ordinal Regression is classification where the class labels are ranked and the error grows with the distance between the correct label and the predicted label \cite{Gutierrez2016}. These errors need not necessarily be linear. Consider for example our case of benign, PUP, and malicious labels. The errors for marking benign as PUP could be 0.1 points, and PUP as malware results in 0.9 units of error. Then marking something benign as malicious would incur 0.1+0.9=1.0 units of error. This type of ordinal regression could be further extended to take into account the severity of certain types of malware. For example, one could instead classify binaries as "benign, PUP, low risk malware, high risk malware". This would naturally require a distinction between risk levels for malware, which could be user dependent. One reasonable distinction between low and high risk malware could be data loss. A machine attached to a botnet would then be low risk, but ransomware would be high risk.  Such alternative labeling schemes are an open area for research and discussion. 

\subsection{Automatic Malware Remediation}

One unit of functionality currently provided by anti-virus products is the removal of found malware. This is a critical function of many anti-virus products, and the time it takes analysts to devise a removal strategy can become part of the lag time between initial malware discovery and updates being sent to user systems. Automating this task, even to only an incremental degree, would help reduce human time spent on this problem and provide relief to users when new malware is found. 

We are not aware of any work that has yet explored the ability for a machine learning system to determine how to remove malware from an already infected system. Given a corpus annotated with the various methods by which malware may attempt to gain persistence on a machine, it would seem plausible to detect these mechanisms and develop a "removal policy", a series of steps to perform that would remove the malware. This would fall into the domain of Reinforcement Learning as a method of learning and executing such policies. 

Multiple malware infections could complicate this task in interesting ways, increasing the challenge involved. The task is also impacted by the operating system version in used, updates installed and available, and other applications installed on the machine. It is unlikely that initial work will address all of these confounding factors at once, but the situations presents a challenging area for AI that could have important practical consequences if successful. One small advantage to the situation, because it is applied only to found malware, arises since it is realistic to use more compute intensive methods for determining a removal policy. This task is unexplored at the moment, and we are not aware of any corpora with such labels. 

\subsection{Malware, Natural Language Processing, and Reports }

A common task for malware analysts is to generate reports of their findings, which are shared with other security processionals to make them aware of new threats, and how to identify the malware, and other valuable information. These documents may be in any format, and may contain a variety of detail labels. Little work has been done on this data, mostly consisting of performing a variety of NLP tasks on the reports themselves 
\cite{lim-etal-2017-malwaretextdb,10.1145/3341161.3343519}. 

In reality, the reports represent one mode of a multi-model data tuple, the textual report of behavior and unique identifies, and the unstructured binary executable(s) that are the subject of the report. A regular occurrence is a new malware family being discovered, and these reports may serve as a source of additional information that could be used to detect novel malware families with limited labeled examples. Other possibilities include work in generating these reports from the malware itself, following inspiration from the automated statistician work \cite{Nguyen2019,steinruecken2019a,Hwang2016,grosse2012a,10.5555/2893873.2894066}. This could aid in both developing/adapting models more quickly, as well as disseminating information faster. A variety of possibilities exist at this unique intersection of malware detection and NLP that have yet to be explored. 

\section{Conclusion} \label{sec:conclusion}

We have given an overview of the machine learning methods currently used for the task of malware classification, as well as the features that are used to drive them. At the same time a number of challenges specific to malware have impeded progress and make objective evaluations difficult. These challenges permeate all stages of the machine learning process and touch upon a vast range of computer science sub-domains, requiring many solutions at different levels of the problem. While we have touched upon some avenues of research that have not yet been addressed by the current literature, many of the challenges remaining will require significant engineering effort and community cooperation and action to overcome. This is especially true of the data collection and processing stages, which are vital to all other stages of the machine learning process, due to the number of unique challenges and costs associated with processing binary data. 

\bibliographystyle{IEEEtranN}
\bibliography{Mendeley}

\end{document}